\documentclass[a4paper,fleqn,usenatbib,useAMS]{mnras}

\usepackage[T1]{fontenc}
\usepackage{ae,aecompl}

\usepackage{graphicx}	
\usepackage{amsmath}	
\usepackage{amssymb}	
\usepackage{color}
\usepackage{multicol}
\usepackage{longtable}
\usepackage{rotating}
\usepackage{pdflscape}
\usepackage{natbib}
\usepackage{enumitem}
\usepackage{appendix}
\usepackage{url}
\usepackage{float}
\usepackage{tabularx}
\usepackage{booktabs}
\usepackage[caption = false]{subfig}

\title[Stellar and dynamical masses of BCGs II]{Dynamical masses of brightest cluster galaxies II: constraints on the stellar IMF}

\author[Loubser et al.]{S. I. Loubser$^{1}$\thanks{E-mail:Ilani.Loubser@nwu.ac.za (SIL)}, H. Hoekstra$^{2}$, A. Babul$^{3}$, Y. M. Bah\'{e}$^{2}$, M. Donahue$^{4}$\\
$^{1}$Centre for Space Research, North-West University, Potchefstroom 2520, South Africa\\
$^{2}$Leiden Observatory, Leiden University, PO Box 9513, 2300 RA, Leiden, The Netherlands\\
$^{3}$Department of Physics and Astronomy, University of Victoria, Victoria, BC, V8W 2Y2, Canada\\
$^{4}$Michigan State University, Physics $\&$ Astronomy Dept., East Lansing, MI 48824-2320, USA}

\date{Accepted 2020 November 10. Received 2020 November 03; in original form 2020 July 06}

\pubyear{2020}

\begin{document}
\label{firstpage}
\pagerange{\pageref{firstpage}--\pageref{lastpage}}
\maketitle

\begin{abstract}
We use stellar and dynamical mass profiles, combined with a stellar population analysis, of 32 brightest cluster galaxies (BCGs) at redshifts of 0.05 $\leq z \leq$ 0.30, to place constraints on their stellar Initial Mass Function (IMF). We measure the spatially-resolved stellar population properties of the BCGs, and use it to derive their stellar mass-to-light ratios ($\Upsilon_{\star \rm POP}$). We find young stellar populations ($<$200 Myr) in the centres of 22 per cent of the sample, and constant $\Upsilon_{\star \rm POP}$ within 15 kpc for 60 per cent of the sample. We further use the stellar mass-to-light ratio from the dynamical mass profiles of the BCGs ($\Upsilon_{\star \rm DYN}$), modelled using a Multi-Gaussian Expansion (MGE) and Jeans Anisotropic Method (JAM), with the dark matter contribution explicitly constrained from weak gravitational lensing measurements. We directly compare the stellar mass-to-light ratios derived from the two independent methods, $\Upsilon_{\star \rm POP}$ (assuming some IMF) to $\Upsilon_{\star \rm DYN}$ for the subsample of BCGs with no young stellar populations and constant $\Upsilon_{\star \rm POP}$. We find that for the majority of these BCGs, a Salpeter (or even more bottom-heavy) IMF is needed to reconcile the stellar population and dynamical modelling results although for a small number of BCGs, a Kroupa (or even lighter) IMF is preferred. For those BCGs better fit with a Salpeter IMF, we find that the mass-excess factor against velocity dispersion falls on an extrapolation (towards higher masses) of known literature correlations. We conclude that there is substantial scatter in the IMF amongst the highest-mass galaxies.
\end{abstract}





\section{Introduction}
\label{introduction}


One of the biggest unknowns of stellar population and mass studies of galaxies is the stellar Initial Mass Function (IMF), the distribution of stellar masses at birth. Limited information regarding the shape and properties of the IMF can be constructed from first principles, e.g.\ using analytic models \citep{Hopkins2013} or hydrodynamical simulations \citep{Krumholz2016}, and must mainly be empirically determined. First thought to be universal \citep{Bastian2010}, the last decade has seen increasing evidence suggesting that the IMF varies between galaxies, as well as within individual galaxies (e.g. \citealt{Conroy2012}, \citealt{Cappellari2013}, \citealt{Ferreras2013}, \citealt{MartinNavarro2015a}, \citealt{Lyubenova2016}, \citealt{Davis2017}). 

Since some stellar absorption line features are stronger in dwarf than giant stars \citep{Spinrad1971}, one method to constrain the IMF (from integrated stellar light) is stellar population synthesis and direct measurements of IMF-sensitive spectral absorption features \citep{Conroy2012, Spiniello2012, LaBarbera2015, Rosani2018, LaBarbera2019}. Another method is to dynamically measure the stellar mass-to-light ratio ($\Upsilon_{\star}$), by assuming the system is dynamically relaxed and by making assumptions about the contribution of the dark matter and supermassive black hole to the gravitational potential of the galaxy \citep{Thomas2011D, Tortora2013, Cappellari2013}. A third independent method to constrain the IMF is through gravitational lensing \citep{Auger2010, Treu2010, Dutton2013, Smith2015, Collier2018}. 

Studies utilising a combination of these independent techniques have found evidence for an IMF that is `heavier', i.e. with larger stellar mass-to-light ratios than those predicted by Chabrier \citep{Chabrier2003} or Kroupa \citep{Kroupa2001} IMFs, in the most massive early-type galaxies \citep{Treu2010, Cappellari2012, Conroy2012, LaBarbera2013, Tortora2013}. The excess mass appears to increase with the stellar velocity dispersion (a proxy for the mass, or similarly with correlated properties e.g. metallicity or [$\alpha$/Fe]-enhancements) of the galaxy \citep{Conroy2012, LaBarbera2015, Rosani2018}. This qualitative agreement between different techniques has lent confidence to claims of a non-universal, heavy IMF for massive early-type galaxies. 

However, \citet{Smith2015} investigated early-type galaxy strong lenses from the SINFONI Nearby Elliptical Lens Locator Survey (SNELLS), and based on their lensing masses, concluded that these galaxies have a stellar IMF consistent with that of the Milky Way, and not the bottom-heavy IMF typically reported for massive early-type galaxies. \citet{Leier2016} also exclude a very bottom-heavy IMF in their study of 18 massive early-type galaxies from the Sloan Lenses ACS Survey (SLACS). \citet{Smith2017} further present dynamical modelling of the BCG in Abell 1201. By using a combination of lensing and stellar dynamics, and by imposing a standard NFW dark matter density profile \citep{Navarro1996}, they recover a stellar mass-to-light ratio that is consistent with a Milky-Way-like IMF.


\citet{Smith2014} pointed out that the IMF constraints obtained for individual galaxies using different techniques do not always agree, and that different studies do not agree on the underlying principal galaxy property tied to the IMF trends (see also \citealt{Newman2017}). However, the study by \citet{Lyubenova2016}, which constrained the IMF of 27 early-type galaxies using stellar populations and dynamics from CALIFA data, suggested the opposite, i.e.\ that there are no disagreements on a case-by-case basis. 


In this paper, we focus on the most massive early-type galaxies. We use a large sample of 32 nearby ($z < 0.3$) brightest cluster galaxies (BCGs), from the Multi-Epoch Nearby Cluster Survey (MENeaCS) and Canadian Cluster Comparison Project (CCCP) cluster samples. The BCG sample spans $M_{K}$ = --25.7 to --27.8 mag, with host cluster halo masses $M_{500}$ between $2.0\times10^{14}$ and $1.5\times10^{15}$ M$_{\sun}$ \citep{Herbonnet2019}. In an accompanying paper \citep{Loubser2020a}, we have modelled the stellar and dynamical mass profiles of 25 out the 32 BCGs (excluding the seven for which we found multiple nuclei or significant substructure in their nuclei) using an adapted Multi-Gaussian Expansion (MGE, \citealt{Monnet1992, Emsellem1994, Cappellari2002}) technique and Jeans Anisotropic Method (JAM, \citealt{Cappellari2006, Cappellari2008}) for an axisymmetric case, deriving the stellar mass-to-light ratio ($\Upsilon_{\star \rm DYN}$), and stellar velocity anisotropy ($\beta_{z}$). 

One major uncertainty in stellar dynamical modelling is the contribution of the dark matter halo to the potential of the galaxies. We were able to estimate the dark matter mass ($M_{\rm DM}$) from weak lensing observations \citep{Herbonnet2019} for 23 of the 25 MENeaCS and CCCP clusters, and thereby limit the number of free parameters in the dynamical models. This, together with our large sample size, is a considerable improvement on previous studies modelling the stellar dynamics of BCGs. 

Here, we present the spatially-resolved stellar population properties of (all 32 of) the BCGs, and use it to estimate their stellar mass-to-light ratios ($\Upsilon_{\star \rm POP}$), under the assumption of both a Salpeter and a Kroupa IMF. We then compare the stellar mass-to-light ratios derived from the two independent methods ($\Upsilon_{\star \rm POP}$ with $\Upsilon_{\star \rm DYN}$) for a subsample of BCGs (i.e.\ those with no young stellar populations and a non-variable $\Upsilon_{\star \rm POP}$ over our radial range) and use it to constrain their stellar IMF. 


We use $H_{0}$ = 73 km s$^{-1}$ Mpc$^{-1}$, $\Omega_{\rm matter}$ = 0.27, $\Omega_{\rm vacuum}$ = 0.73 throughout. We refer to velocity dispersion as $\sigma$, to rotational velocity as $V$, and to $\sqrt{V^{2} + \sigma^{2}}$ as the second moment of velocity ($\nu_{\rm RMS}$). We also derive the stellar mass-to-light ratios, $\Upsilon_{\star \rm DYN}$ and $\Upsilon_{\star \rm POP}$, in the rest-frame $r$-band. In Section \ref{data} we briefly summarise our data as well as the dynamical modelling procedure used to derive the stellar mass-to-light ratios ($\Upsilon_{\star \rm DYN}$) in \citet{Loubser2020a}. In Section \ref{StelPops} we present the spatially-resolved stellar population properties, and the predicted stellar mass-to-light ratios ($\Upsilon_{\star \rm POP}$), assuming a Salpeter IMF \citep{Salpeter1955}. We then also fit the stellar populations using a Kroupa IMF \citep{Kroupa2001}, and discuss the constraints on the IMF in Section \ref{Constraints_IMF}. We correlate the mass excess factor ($\alpha =  \log_{10} ( \Upsilon_{\star \rm DYN}) -  \log_{10} ( \Upsilon_{\star \rm POP})$) with other galaxy properties in Section \ref{corr}, and summarise our conclusions in Section \ref{conclusions}. 

\section{Data and stellar mass-to-light ratios from dynamics ($\Upsilon_{\star \rm DYN}$)}
\label{data}

We use spatially-resolved, long-slit spectroscopy for 14 MENeaCS and 18 CCCP BCGs, observed on the Gemini North and South telescopes, and \textit{r}-band imaging observed on the Canada-France-Hawaii telescope (CFHT). We also use host cluster properties derived from \textit{Chandra/XMM-Newton} X-ray data, and cluster masses measured through weak lensing \citep{Mahdavi2013, Hoekstra2015, Herbonnet2019}. 

The stellar population analysis and star formation histories of the CCCP BCGs are presented in \citet{Loubser2016}, but since we present the stellar population analysis for the MENeaCS BCGs here for the first time, we briefly summarise the relevant properties of the long-slit spectroscopic data used for the MENeaCS analysis. The 14 MENeaCS BCGs were observed using the Gemini Multi-Object Spectrograph (GMOS) detector during two semesters (GS2009A, GN2009A, GN2009B). The instrumental configuration consisted of the B600 grating at a central wavelength of 4600 \AA{}, and a slit width of 1 arcsec. We used 2$\times$2 binning, corresponding to an instrumental resolution of 71 km s$^{-1}$. The spatial apertures (i.e.\ 0 -- 5 kpc, and 5 -- 15 kpc) were chosen to ensure sufficient S/N for reasonable errors in the stellar population analysis for all BCGs, with the S/N of the MENeaCS spectra generally higher than that of the CCCP spectra presented in \citet{Loubser2016}.

In \citet{Loubser2020a}, we present the Gauss-Hermite higher order velocity moments $h_{3}$ and $h_{4}$ for the BCGs, and find that the central measurements of $h_{4}$ are positive for all our BCGs. We then model the stellar and dynamical mass profiles of 25 out of the 32 BCGs (excluding the seven for which we found multiple nuclei or significant substructure in their nuclei), using an adapted Multi-Gaussian Expansion (MGE) and Jeans Anisotropic Method (JAM) for an axisymmetric case (for both cylindrically- and spherically aligned models), deriving the stellar mass-to-light ratio ($\Upsilon_{\star \rm DYN}$), and anisotropy ($\beta_{z}$)\footnote{For the cylindrically-aligned axisymmetric models, the velocity anisotropy is defined as $\beta_{z} = 1 - (\sigma^{2}_{z} / \sigma^{2}_{R})$, and for the spherically-aligned axisymmetric models as $\beta = 1 - (\sigma^{2}_{\theta} / \sigma^{2}_{R}$). We show in Appendix \ref{spherical} that the choice of alignment does not influence our conclusions.}, where the dark matter mass was constrained from weak lensing results. Our fits to the observed kinematics are restricted to the galaxy centre, where the stellar component is the dominant contributor to the mass ($<$20 kpc).

Our dynamical modelling revealed that the stellar anisotropy and velocity dispersion profile slope ($\eta$, from \citealt{Loubser2018}) are strongly correlated. The BCGs with rising velocity dispersion profiles showed tangential anisotropy parameters, whereas the BCGs with decreasing velocity dispersion profiles showed radial anisotropy parameters. The positive velocity dispersion gradients (i.e. profiles rising from the centre of the BCG), can also arise from a significant contribution from the intracluster light (ICL). For a small subset of BCGs with positive velocity dispersion gradients, a variable $\Upsilon_{\star \rm DYN}$ can also explain the profile shape, instead of tangential anisotropy or an ICL contribution.


\section{Stellar population analysis}
\label{StelPops}

In \citet{Loubser2016}, we identified plausible star formation histories for the 18 CCCP BCGs for which we have long-slit spectra by fitting simple stellar populations (SSPs) and composite populations consisting of a young stellar component and an intermediate or old stellar component. In this Section, we repeat the analysis for the 14 MENeaCS BCGs, located at lower redshifts. We use the University of Lyon Spectroscopic analysis Software (ULySS, see \citealt{Koleva2009, Groenewald2014, Loubser2016}) to fit stellar population models to the BCG spectra, taking into account the internal kinematics of the galaxies. We use the \citet{Vazdekis2010} stellar population models, based on the MILES library \citep{Sanchez2006}, first for a Salpeter IMF \citep{Salpeter1955}, and then for a Kroupa IMF \citep{Kroupa2001}. Any emission lines are masked, and the rest of the spectrum is then used to determine the best-fitting stellar population model. We identify both the best-fitting spectrum resulting from a single SSP, as well as the best-fitting composite stellar model comprising of a young stellar population superposed on an intermediate or old stellar population, and determine which best describe the observed spectrum\footnote{We refer the reader to \citet{Loubser2016} where we do various tests to show that we can robustly detect young stellar populations in the BCG spectra.}. We further describe the BCGs with young stellar components, and the stellar mass-to-light ratios ($\Upsilon_{\star \rm POP}$) below.

To assess possible systematic errors in our stellar population analysis, and the possible effect on $\Upsilon_{\star \rm POP}$, we also use the full spectral fitting code FIREFLY (Fitting IteRativEly For Likelihood analYsis) as described in \citet{Wilkinson2017}, with the MaStar stellar population models \citep{Maraston2020}, in Section \ref{systematic_errors}, and show that the modelling approach does not affect our main conclusions.

\subsection{BCGs containing young stellar population components}

We detected prominent young ($\sim$200 Myr) stellar populations in four of the 18 CCCP BCGs (see \citealt{Loubser2016}). We repeat the analysis on the MENeaCS BCGs, with the same spatial bins as used in \citet{Loubser2016}, i.e.\ 0 -- 5 kpc, and 5 -- 15 kpc, and find three of the 14 MENeaCS BCGs (Abell 780, 1795 and 2055) show very young stellar populations ($<$200 Myr) in their inner apertures. Therefore, in total 7/32 (22 per cent) of the BCG sample show a young stellar population component. Two of these BCGs, Abell 780 and Abell 2055, have intermediate/old SSP-equivalent ages in their outer bins (and thus relatively large SSP-equivalent age gradients), whereas Abell 1795 has a fairly young ($\sim$2 Gyr) SSP-equivalent age in its outer aperture. Only two MENeaCS BCG, Abell 1991 and 2319, have very old ($>$10 Gyrs) stellar populations in both the inner and outer apertures. The nine remaining MENeaCS BCGs all present intermediate stellar populations in their inner and outer bins. See \citet{Loubser2016} for the complete discussion on intermediate SSP-equivalent ages in BCGs. The stellar population results for the MENeaCS subsample are presented in Table \ref{ages} (here), and the CCCP subsample in \citet{Loubser2016} (their Table 2).

\begin{table*}
\caption{Stellar population properties of the MENeaCS BCGs. The young (1) and old (2) stellar population components are indicated for the three BCGs that contain young stellar populations. The four BCGs marked with `$\star$' have emission lines in their spectra. The last column shows the colour of the core \citep{Bildfell2008}.}
\label{ages}
\begin{tabular}{l c c r r r r c}
\hline
Name & Aperture & Component & Age  & [Fe/H]  & Luminosity fraction & Mass fraction & Classification\\
&  &  & $\rmn{(Myr)}$ & (dex) &   &  & \\
\hline
\multicolumn{8}{c}{Young stellar components}\\
Abell 780$^\star$ & inner & 1 & 100 $\pm$ 50 & --1.46 $\pm$ 0.16 & 47 & 1 & blue\\
& inner & 2 & 10310 $\pm$ 960 & 0.20 $\pm$ 0.05 & 53 & 99 & \\
&  outer & SSP & 15140 $\pm$ 5280 & --0.12 $\pm$ 0.16 & 100 & 100 & \\
Abell 1795$^\star$ & inner & 1 & 120 $\pm$ 30 & --1.43 $\pm$ 0.14 & 27 & 1 & blue\\
& inner & 2 & 6660 $\pm$ 2790 & 0.19 $\pm$ 0.01 & 73 & 99 & \\
&  outer & SSP & 2210 $\pm$ 240 & 0.09 $\pm$ 0.03 & 100 & 100 & \\
Abell 2055$^\star$ & inner & 1 & 100 $\pm$ 50 & --0.80 $\pm$ 0.90 & 14 & $<$1 & blue\\
& inner & 2 & 11300 $\pm$ 3150 & --1.30 $\pm$ 0.40 & 86 & $>$99 & \\
&  outer & SSP & 4090 $\pm$ 980 & --1.27 $\pm$ 0.03 & 100 & 100 & \\
\hline
\multicolumn{8}{c}{Intermediate stellar population}\\
Abell 644 & inner & SSP & 4870 $\pm$ 1160 & 0.19 $\pm$ 0.02 & 100 & 100 & red\\
&  outer & SSP & 5490 $\pm$ 1610 & 0.20 $\pm$ 0.05 & 100 & 100 & \\
Abell 646$^\star$ & inner & SSP & 3680 $\pm$ 510 & 0.07 $\pm$ 0.07 & 100 & 100 & blue\\
&  outer & SSP & 5780 $\pm$ 2150 & 0.15 $\pm$ 0.07 & 100 & 100 & \\
Abell 754 & inner & SSP & 7730 $\pm$ 300 & 0.20 $\pm$ 0.05 & 100 & 100 & red\\
&  outer & SSP & 7540 $\pm$ 270 & 0.18 $\pm$ 0.02 & 100 & 100 & \\
Abell 990 & inner & SSP & 4280 $\pm$ 770 & 0.19 $\pm$ 0.05 & 100 & 100 & red\\
&  outer & SSP & 4610 $\pm$ 500 & 0.20 $\pm$ 0.05 & 100 & 100 & \\
Abell 1650 & inner & SSP & 6640 $\pm$ 970 & 0.20 $\pm$ 0.01 & 100 & 100 & red\\
&  outer & SSP & 5550 $\pm$ 370 & 0.20 $\pm$ 0.05 & 100 & 100 & \\
Abell 2029 & inner & SSP & 7360 $\pm$ 830 & 0.20 $\pm$ 0.05 & 100 & 100 & red\\
&  outer & SSP & 7890 $\pm$ 5660 & 0.16 $\pm$ 0.17 & 100 & 100 & \\
Abell 2050 & inner & SSP & 6300 $\pm$ 1750 & 0.20 $\pm$ 0.05 & 100 & 100 & red\\
&  outer & SSP & 3980 $\pm$ 290 & 0.20 $\pm$ 0.05 & 100 & 100 & \\
Abell 2142 & inner & SSP & 7040 $\pm$ 1270 & 0.19 $\pm$ 0.02 & 100 & 100 & red\\
&  outer & SSP & 17430 $\pm$ 710 & --0.28 $\pm$ 0.05 & 100 & 100 & \\
Abell 2420 & inner & SSP & 7650 $\pm$ 360 & 0.20 $\pm$ 0.05 & 100 & 100 & red\\
&  outer & SSP & 5920 $\pm$ 1330 & 0.20 $\pm$ 0.05 & 100 & 100 & \\
\hline
\multicolumn{8}{c}{Old stellar population}\\
Abell 1991 & inner & SSP & 13260 $\pm$ 5860 & --0.05 $\pm$ 0.19 & 100 & 100 & red\\
&  outer & SSP & 10310 $\pm$ 5290 & 0.08 $\pm$ 0.02 & 100 & 100 & \\
Abell 2319 & inner & SSP & 17520 $\pm$ 450 & 0.13 $\pm$ 0.05 & 100 & 100 & red\\
& outer  & SSP & 17780 $\pm$ 200 & 0.09 $\pm$ 0.05 & 100 & 100 &\\
\hline
\end{tabular}
\end{table*}

\subsection{Stellar mass-to-light ratios from stellar populations ($\Upsilon_{\star \rm POP}$)}
\label{MtoL}

In addition to identifying and constraining young stellar populations, we use the stellar population results to determine the stellar mass-to-light ratios ($\Upsilon_{\star \rm POP}$), which depend sensitively on the stellar IMF. The BCG stellar mass-to-light ratios from dynamics ($\Upsilon_{\star \rm DYN}$), which are treated as a (constant) free parameter in the dynamical mass models, can be directly compared to $\Upsilon_{\star \rm POP}$ (for different IMFs) to place constraints on the IMF.

We use the ages and metallicities derived (Table \ref{ages}), and photometric predictions from the Vazdekis/MILES \citep{Sanchez2006, Vazdekis2015} models with the \citet{Girardi2000} isochrones with solar [$\alpha$/Fe]-enhancement, to determine $\Upsilon_{\star \rm POP}$ for the $r$-filter in the inner (0 -- 5 kpc) and outer (5 -- 15 kpc) apertures. For the BCGs with younger components, we derive a light-weighted average $\Upsilon_{\star \rm POP}$ of the composite stellar populations. If the $\Upsilon_{\star \rm POP}$ derived in the inner and outer bins agree within their 1$\sigma$ errors (as propagated from the errors on the ages and metallicities), then we consider the $\Upsilon_{\star \rm POP}$ to remain constant with radius within this spatial range in the BCG. 

We present these results (for a Salpeter IMF) in Table \ref{properties} for all 32 BCGs. We find that 19/32 (60 per cent) of the BCGs have constant $\Upsilon_{\star \rm POP}$ over this radial range (0 -- 15 kpc). Since the difference between the stellar population properties derived for the inner and outer apertures are relative, whether $\Upsilon_{\star \rm POP}$ is relatively insensitive to the systematic uncertainties connected to the stellar population models or the assumed IMF.  

\begin{table}
\caption{Averaged stellar mass-to-light ratios ($\Upsilon_{\star \rm POP}$) of the CCCP and MENeaCS BCGs (for a Salpeter IMF). The BCGs with constant $\Upsilon_{\star \rm POP}$ are indicated with a Y in column four. The cooling time $t_{\rm c,0}$ is measured at a radius of 20 kpc from \textit{Chandra}, and \textit{XMM-Newton} X-ray data by \citet{Mahdavi2013}. The BCGs marked with a `$\star$' satisfy our subsample selection criteria as discussed in Section \ref{15}.}
\label{properties}
\begin{tabular}{l c c c r }
\hline
Name & $z$   &   $\Upsilon_{\star \rm POP}$ & $\Upsilon_{\star \rm POP}$  & $t_{c,0}$  \\
            &                           &                               & constant            &                      (Gyr)      \\             
\hline
\multicolumn{5}{c}{MENeaCS}\\
Abell 780 & 0.054 &		3.88 $\pm$ 0.46&   & 0.59 $\pm$ 0.00 \\
Abell 754$^{\star}$ & 0.054 &	 	4.44 $\pm$ 0.23&	Y  &   6.84 $\pm$ 0.25 \\ 
Abell 2319 & 0.056 &	7.20 $\pm$ 0.31&	Y  &  5.55 $\pm$ 0.15 \\
Abell 1991$^{\star}$ & 0.059 &	5.20 $\pm$ 0.98&	Y   &     0.64 $\pm$ 0.01 \\
Abell 1795 & 0.063 &		2.30 $\pm$ 0.12&	   &   0.69 $\pm$ 0.01  \\
Abell 644 & 0.070  &	 	3.40 $\pm$ 0.39 &	Y  &      2.98 $\pm$ 0.18 \\
Abell 2029$^{\star}$ & 0.077  &		4.08 $\pm$ 1.04 &	Y   &     0.71 $\pm$ 0.01  \\
Abell 1650$^{\star}$ & 0.084  &			3.74 $\pm$ 0.24&	Y   &     1.58 $\pm$ 0.04  \\
Abell 2420$^{\star}$ & 0.085  &	 	4.13 $\pm$ 0.36 &	Y   &       10.37 $\pm$ 3.89 \\
Abell 2142 & 0.091  &		5.15 $\pm$ 0.50 &	   &        1.32 $\pm$ 0.05 \\
Abell 2055 & 0.102  &		 1.96 $\pm$ 0.21 &	    &      79.71 $\pm$ 38.00  \\
Abell 2050$^{\star}$ & 0.118  &		3.09 $\pm$ 0.41&	Y  &        3.76 $\pm$ 0.19 \\
Abell 646$^{\star}$ & 0.129  &		 	2.76 $\pm$ 0.45&	Y   &        --   \\
Abell 990 & 0.144  &			2.93 $\pm$ 0.21&	Y   &       --  \\
\hline
\multicolumn{5}{c}{CCCP}\\
Abell 2104 & 0.153  &		3.31 $\pm$ 0.15&	  &       5.52 $\pm$ 1.00  \\
Abell 2259$^{\star}$ & 0.164 &	3.00 $\pm$ 0.15&	Y  &       3.71 $\pm$ 0.77  \\
Abell 586 & 0.171  &		2.88 $\pm$ 0.21&	Y  &       2.86 $\pm$ 0.45 \\
MS 0906+11 & 0.174  &	 3.25 $\pm$ 0.14&	  &       2.88 $\pm$ 0.54 \\
Abell 1689$^{\star}$ & 0.183 &		4.93 $\pm$ 0.17&	Y  &        1.19 $\pm$ 0.05\\
MS 0440+02 & 0.187   &		3.49 $\pm$ 0.17&	  &        0.73 $\pm$ 0.11 \\ 
Abell 383 & 0.190   &			2.62 $\pm$ 0.18&	Y   &       0.41 $\pm$ 0.02 \\
Abell 963 & 0.206  &		4.83 $\pm$ 0.14&	  &      1.32 $\pm$ 0.07  \\
Abell 1763$^{\star}$ & 0.223   &			4.39 $\pm$ 0.20&	Y   &       10.61 $\pm$ 1.38\\
Abell 1942$^{\star}$ & 0.224    &		2.86 $\pm$ 0.12&	Y  &       6.26 $\pm$ 2.34 \\
Abell 2261$^{\star}$ & 0.224  &		2.98 $\pm$ 0.24&	Y   &      1.14 $\pm$ 0.14 \\
Abell 2390 & 0.228  &	2.14 $\pm$ 0.13&	  &      0.58 $\pm$ 0.01 \\
Abell 267$^{\star}$ & 0.231  &	2.80 $\pm$ 0.04&	Y  &        4.15 $\pm$ 0.51 \\
Abell 1835 & 0.253   &1.73 $\pm$ 0.09&	  &        0.29 $\pm$ 0.00 \\
Abell 68$^{\star}$ & 0.255   &	 	3.00 $\pm$ 0.20&	Y   &      3.57 $\pm$ 0.73 \\
MS 1455+22 & 0.258   &	 	1.67 $\pm$ 0.10&	   &        0.40 $\pm$ 0.01 \\
Abell 611 & 0.288   & 	3.81 $\pm$ 0.24 &	  &       1.28 $\pm$ 0.21  \\
Abell 2537 & 0.295 &	3.99 $\pm$ 0.26&	  &      2.05 $\pm$ 0.39  \\
\hline
\end{tabular}
\end{table}

\begin{figure}
   \includegraphics[scale=0.34]{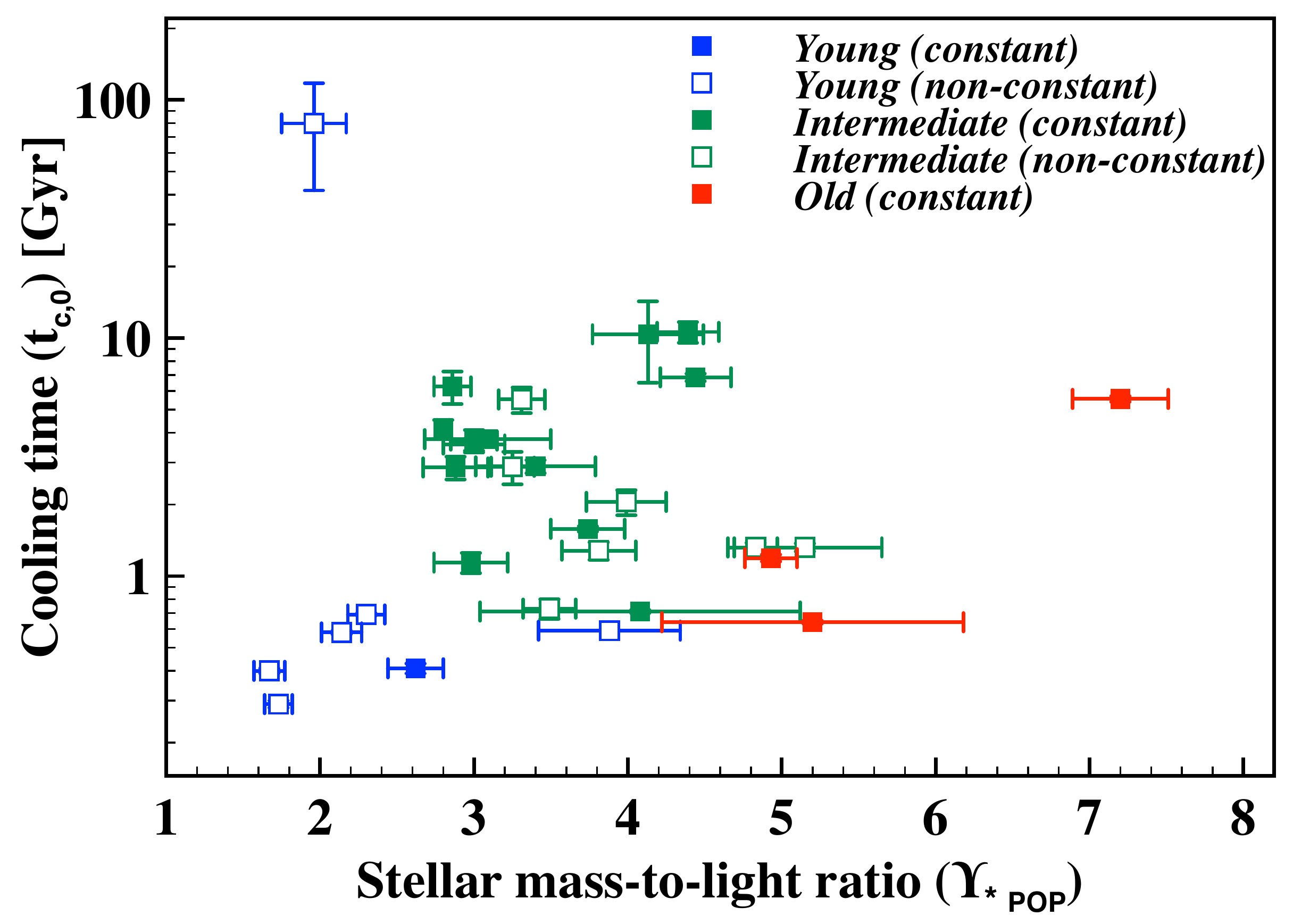}
   \caption{The stellar mass-to-light ratio ($\Upsilon_{\star \rm POP}$) against central cooling time, $t_{\rm c,0}$ (in Gyr). Blue indicates BCGs with young components (see top panel of Table \ref{ages}) present, green intermediate aged BCGs and red are old (10 Gyr or older in both the inner and outer bin). The empty symbols are those galaxies where $\Upsilon_{\star \rm POP}$ is not constant between the inner and outer apertures. The central cooling time $t_{\rm c,0}$ is measured at a radius of 20 kpc from \textit{Chandra, XMM-Newton} X-ray data by \citet{Mahdavi2013} and given in Table \ref{properties}. The outlier in the top left corner is the BCG in Abell 2055 (a BL Lac), and described in Section \ref{disresults}.}
\label{fig:ML}
\end{figure}

\subsection{Discussion of stellar population modelling results}
\label{disresults}

We plot the stellar mass-to-light ratio ($\Upsilon_{\star \rm POP}$) against central cooling time ($t_{\rm c,0}$) in Figure \ref{fig:ML}. Blue symbols indicate BCGs with young components, green intermediate aged and red old stellar populations. The solid symbols are those galaxies for which the $\Upsilon_{\star \rm POP}$ is constant between the inner and outer apertures (within the errors), and the empty symbols non-constant (primarily driven by age gradients between the inner and outer apertures). From Figure \ref{fig:ML} it follows, similar to our conclusions in \citet{Loubser2016}, that the BCGs with young stellar populations are located in host clusters with short cooling times, with the exception of the BCG in Abell 2055. The BCG in Abell 2055 (top left corner in Figure \ref{fig:ML}) shows optical emission lines in its spectrum, a blue core \citep{Bildfell2008}, and a young stellar component, however the host cluster has the longest cooling time of all the clusters considered here. This BCG hosts a BL Lac point source that is the dominant contributor to the observed emission \citep{Green2017}. 

Unlike the BCGs in Abell 780, 1795 and 2055, which show young components as well as emission lines in their spectra, the BCG in Abell 646 shows some emission lines in its spectrum, but a stellar component in the inner aperture of age $\sim$3.7 Gyr. It also has a blue core as determined by \citet{Bildfell2008}. There is no significant difference between the $\chi^{2}$-values for the stellar population fits in the inner aperture for an SSP and a two component composite stellar population fit, so we default to an SSP fit (as described in \citealt{Loubser2016}). This is thus an intermediate component galaxy, but with a stellar population component (in the inner bin) on the limit of where a small fraction of young stars cannot be confidently detected (as described in detail in \citealt{Loubser2016}).

We do not find any strong correlations between stellar mass-to-light ratio ($\Upsilon_{\star \rm POP}$) and $K$-band luminosity, central velocity dispersion ($\sigma_{0}$), or $M_{500}$ (not shown here). There are also no clear correlations between the stellar population properties and the stellar velocity anisotropy or velocity dispersion slope \citep{Loubser2018, Loubser2020a}.

Lastly, it is worth noting that most (but not all) BCGs with young stellar components have non-constant $\Upsilon_{\star \rm POP}$. One of the BCGs with a young stellar population component (Abell 383) has constant $\Upsilon_{\star \rm POP}$ (i.e.\ a younger population distributed through the inner and outer apertures).


\section{Comparing $\Upsilon_{\star \rm POP}$ to $\Upsilon_{\star \rm DYN}$ and constraints on the IMF}
\label{Constraints_IMF}
\label{15}

For a given stellar population age and metallicity, the value of $\Upsilon_{\star \rm POP}$ depends strongly on the IMF, since an excess of low-mass stars (or stellar remnants) contributes to the mass without significantly contributing to the luminosity. To compare to $\Upsilon_{\star \rm DYN}$, we use the derived $\Upsilon_{\star \rm POP}$ (averaged over both bins, 0 -- 5 kpc and 5 -- 15 kpc), first for a Salpeter IMF \citep{Salpeter1955}, and then for a Kroupa IMF \citep{Kroupa2001} in the $r$-band (as described in Section \ref{MtoL})\footnote{We repeat this comparison between $\Upsilon_{\star \rm POP}$ and $\Upsilon_{\star \rm DYN}$, but using only the central 5 kpc aperture in Appendix \ref{inner5kpc}.}. We therefore use two different functional forms of the IMF, namely a single (Salpeter) and double (Kroupa) power law. The Salpeter IMF is equivalent to a single power law IMF slope of $\Gamma=1.3$, whereas a Kroupa IMF is indistinguishable from a double power law IMF (two power laws joined by a spline) with a slope of $\Gamma_{b}=1.3$ above 0.6 M$_{\sun}$ and tapered for masses below 0.5 M$_{\sun}$. Both IMFs have a lower and upper mass cutoff of 0.1 M$_{\sun}$ and 100 M$_{\sun}$, respectively.

To determine the $\Upsilon_{\star \rm POP}$ values for the Kroupa IMF, we re-fit the stellar population parameters, similar to the analysis in Section \ref{StelPops}, assuming a Kroupa IMF. We propagate the 1$\sigma$ errors from the stellar population parameters to estimate errors on the resulting $\Upsilon_{\star \rm POP}$. We discuss possible systematic errors on $\Upsilon_{\star \rm POP}$ in Section \ref{systematic_errors}. 

From our dynamical mass models, we use the best-fitting value for the $\Upsilon_{\star \rm DYN}$ parameter (using the stellar, central and dark matter mass components `$\star$ + CEN + DM', from \citet{Loubser2020a}, see summary in Section \ref{data}), where $\Upsilon_{\star \rm DYN}$ was kept constant over our fitting range (<20 kpc)\footnote{We repeat this comparison between $\Upsilon_{\star \rm POP}$ and $\Upsilon_{\star \rm DYN}$, using a parametrised variable $\Upsilon_{\star \rm DYN}$ in our dynamical modelling in Appendix \ref{varML}.}.
The central mass component in our dynamical mass models represents a supermassive black hole (with mass estimated from the M$_{\rmn {BH}} - \sigma$ relation) as described in \citet{Loubser2020a}. To estimate the dark matter mass component, we assume an NFW profile \citep{Navarro1996} and take $M_{\rm DM}$ and $r_{200}$ from weak lensing observations \citep{Herbonnet2019}. We use $M_{\rm DM}=\alpha M_{200}$, where $\alpha = \Omega_{\rm M} /(\Omega_{\rm M} - \Omega_{\rm b})$, and the total mass $M_{200}$ is obtained from weak lensing. We assume the baryon fraction within $r_{200}$ is equal to the cosmological baryon fraction, and therefore $M_{200} \sim 1.2 M_{\rm DM}$, as described in detail in \citet{Loubser2020a}. The dynamical models are free from any assumptions about the stellar populations. We show various robustness tests of our dynamical mass models (e.g. influence of the mass and radius of black hole, influence of the point spread function, and sensitivity of our derived parameters to the dark matter mass distribution and concentration value used for the dark matter halo) in the Appendices in \citet{Loubser2020a}. 

\begin{figure}
\centering
   \includegraphics[scale=0.42]{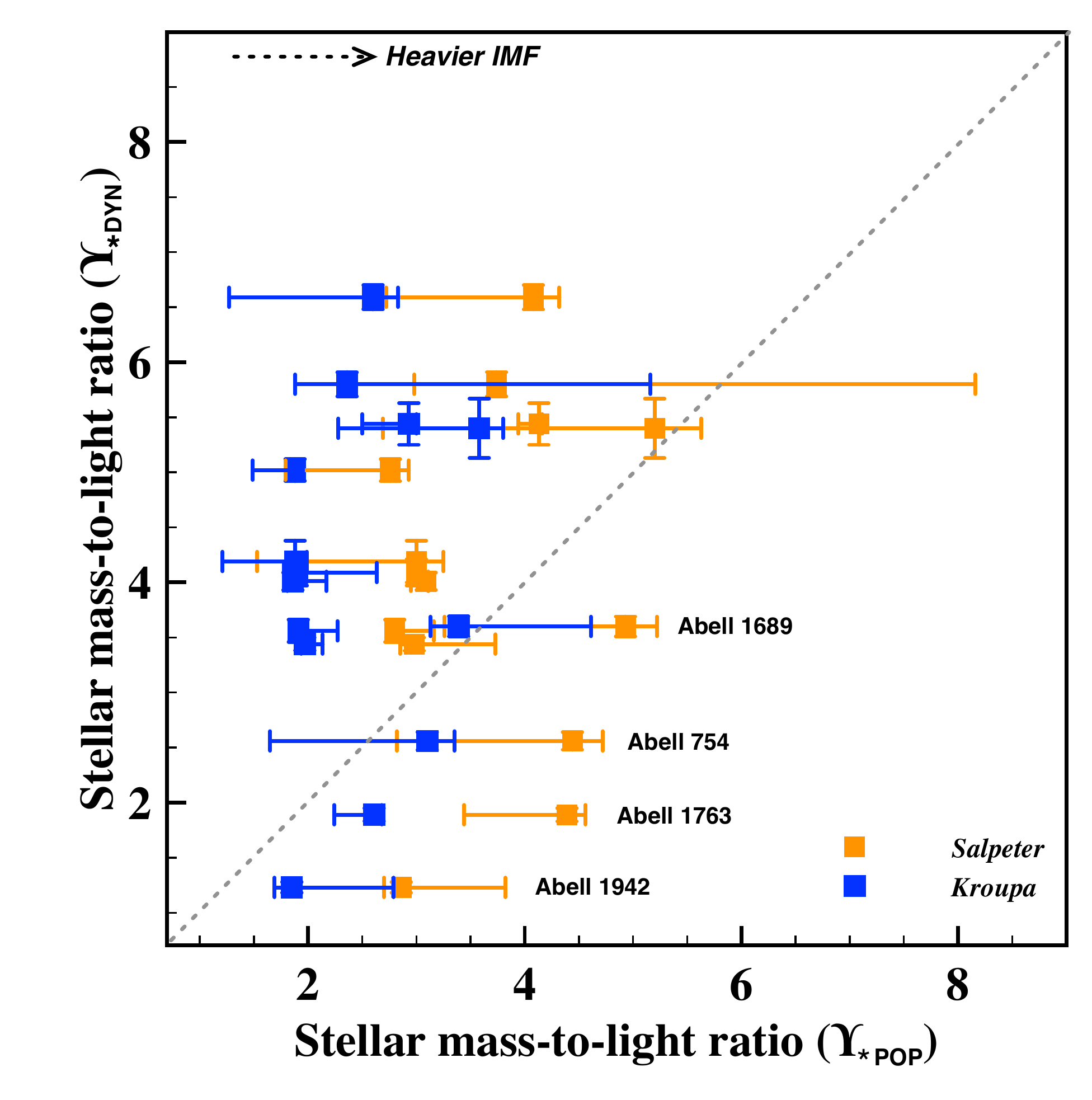}
   \caption{$\Upsilon_{\star \rm DYN}$ (dark matter included) vs $\Upsilon_{\star \rm POP}$ for Salpeter (orange) and Kroupa (blue) IMFs, as discussed in Section \ref{15}. The dotted line indicates the 1-to-1 line, and the arrow in the top left corner indicate that a heavier IMF moves the data points to higher values on the x-axis. Increasing the amount of dark matter in our models move the data points to lower values on the y-axis. Systematic uncertainties on $\Upsilon_{\star \rm DYN}$ can be up to 15 per cent (see Section \ref{systematic_errors} and Appendix \ref{spherical}). The errors on $\Upsilon_{\star \rm POP}$ are the systematic errors as derived and discussed in Section \ref{systematic_errors} and Appendix \ref{MaStar} which give a more realistic representation of the uncertainties than the propagated measurement errors from Table \ref{properties}.}
\label{ML_ML}
\end{figure} 

We aim to eliminate possible uncertainties in the stellar population and dynamical modelling results by comparing $\Upsilon_{\star \rm POP}$ to $\Upsilon_{\star \rm DYN}$ for the BCGs for which we have dynamical models, but eliminate: i) the two BCGs (Abell 963 and 2055) for which we find the best-fitting $\beta_{z}$ in the `$\star$ + CEN + DM' case to indicate extreme tangential anisotropy $\beta_{z}<-1$ (see \citealt{Loubser2020a}); ii) the BCGs where we detect young stellar components; iii) the BCGs where we detect significant age gradients between the inner and the outer stellar population bin (i.e. $\Upsilon_{\star \rm POP}$ is non-constant within 15 kpc). Composite stellar populations would increase $\Upsilon_{\star \rm POP}$ compared to SSPs \citep{Cappellari2006, Trager2008}, although this is much more relevant for low-mass early-type galaxies. Fourteen BCGs, indicated with a `$^{\star}$' in Table \ref{properties}, satisfy these criteria. We plot the comparison between $\Upsilon_{\star \rm POP}$ for a Salpeter (orange symbols) and Kroupa (blue symbols) IMF, and $\Upsilon_{\star \rm DYN}$ in Figure \ref{ML_ML}. We have included the dark matter mass component, estimated from weak lensing measurements for these clusters, in our best-fitting values for $\Upsilon_{\star \rm DYN}$. This decreases $\Upsilon_{\star \rm DYN}$ on average by 8.3 $\pm$ 2.9 per cent over our kinematic range. As expected, a `heavier' IMF shifts the data points towards larger values for $\Upsilon_{\star \rm POP}$ (x-axis). If we would have added even more dark matter in the mass models, it would push $\Upsilon_{\star \rm DYN}$ (y-axis) to lower values (see analysis in \citet{Loubser2020a}).
 
\subsection{Discussion on constraining the IMF}
\label{IMFDiscuss}

From Figure \ref{ML_ML}, we can see that, for a Salpeter IMF, the majority (ten) of the BCGs fall above the 1-to-1 line, but there is a subset of four BCGs below the 1-to-1 line (Abell 754, 1689, 1763 and 1942). The BCGs above the line (with $\Upsilon_{\star \rm DYN} > \Upsilon_{\star \rm POP}$) are better described by the bottom-heavy Salpeter, or an even `heavier', IMF due to a higher fraction of low mass stars. For relatively old stellar populations (as expected for passively-evolving, massive early-type galaxies), only stars with masses below $\sim$1M$_{\sun}$ are present (see \citealt{MartinNavarro2015b, Lyubenova2016}). This limits stellar populations based IMF analysis of massive early-type galaxies to its low-mass end. As seen from Figure \ref{ML_ML}, a Kroupa IMF (which has fewer stars with masses below 0.5M$_{\sun}$ than a Salpeter IMF) describes the four BCGs below the 1-to-1 line better, and is closer to reconciling the stellar mass-to-light ratios measured from stellar population and kinematic properties.

Most studies find a bottom-heavy IMF (e.g. Salpeter) for massive early-type galaxies \citep{Cappellari2013, LaBarbera2015, MartinNavarro2015a}. However, as discussed in Section \ref{introduction}, the SNELLS galaxies yielded lensing masses in strong disagreement with a bottom-heavy IMF for massive early-type galaxies \citep{Smith2017, Newman2017}, instead measuring mass-to-light ratios consistent with a Milky-Way-like IMF (e.g.\ Kroupa), both in low resolution ground based, and in high resolution space-based observations \citep{Collier2018}. However, the SNELLS systems also show other peculiar properties as we discuss further in Section \ref{low_dyn}.

We compare the fraction of dark matter mass (to the total mass), as derived from our dynamical models where the dark matter mass was constrained from weak lensing measurements \citep{Herbonnet2019}, with previous studies. We find an average dark matter fraction of $\sim$8 per cent within 0.38$R_{e}$ as described in \citet{Loubser2020a} (see \citealt{Loubser2020a} for the values for individual clusters). This is consistent with the median dark matter fraction found for massive early-type galaxies in comparable studies (e.g.\ \citealt{Cappellari2013, Lyubenova2016}). Including more mass attributed to dark matter will bring some of the BCGs above the 1-to-1 line closer to the line, but it will move the four BCGs better described by a Kroupa IMF further below the 1-to-1 comparison line. A universal IMF is therefore not only inconsistent with the weak lensing mass measurements, but would also imply high dark matter fractions for some BCGs (in the very centres of the galaxies) and none for others. 

\subsection{Systematic errors on $\Upsilon_{\star \rm DYN}$ and $\Upsilon_{\star \rm POP}$}
\label{systematic_errors}

\subsubsection{$\Upsilon_{\star \rm DYN}$}

BCGs can be classified as oblate, triaxial, or prolate \citep{Krajnovic2018}. We therefore also use the axisymmetric Jeans equations under the assumption of an anisotropic (three-integral) velocity ellipsoid that is aligned with a spherical polar coordinate system \citep{Cappellari2020}. A comparison between the solutions obtained from JAM with spherical polar coordinates and JAM with cylindrical polar coordinates allows for a robust assessment of the best-fitting parameters from the dynamical modelling. For BCGs with decreasing velocity dispersion profiles, $\Upsilon_{\star \rm DYN}$ (spherical coordinates) is up to $\sim$ 15 per cent lower, and for BCGs with rising velocity dispersion profiles, $\Upsilon_{\star \rm DYN}$ (spherical coordinates) is up to $\sim$ 15 per cent higher \citep{Loubser2020a}. We indicate this possible uncertainty in Figure \ref{ML_ML_sph} in Appendix \ref{spherical}, and find that it cannot account for the scatter we observe above or below the 1-to-1 line. 

The IMF may also vary radially within high-mass early-type galaxies, becoming bottom-heavier towards the central regions \citep{VanDokkum2017, Oldham2018, Sarzi2018, Parikh2018, LaBarbera2019}. For M87, \citet{Sarzi2018} found that the IMF drops from a low mass excess in the core to a Milky-Way IMF at 0.4$R_{e}$ (for our BCG sample this corresponds to $\sim$ 15 kpc, on average). \citet{Vaughan2018} find that the IMF for the BCG in the Fornax cluster NGC1399 is heavier than the Milky Way and remains constant out to 0.7$R_{e}$, before it decreases to become marginally consistent with a Milky Way IMF. \citet{Sonnenfeld2019}, for their massive galaxies from the Baryon Oscillation Spectroscopic Survey (BOSS) constant mass (CMASS) sample, find that the region where the IMF is significantly heavier than that of the Milky Way is smaller than the scales probed by the Einstein radius of the lenses in their sample (5 -- 10 kpc). 

As a first test, we also investigate the comparison between $\Upsilon_{\star \rm DYN}$ and $\Upsilon_{\star \rm POP}$ (for a Salpeter and Kroupa IMF), using $\Upsilon_{\star \rm POP}$ for just the central 5 kpc of the BCGs (i.e.\ just the inner aperture, for the same fourteen BCGs). We also expect the dark matter mass component to contribute very little to the total mass in the central 5 kpc, so we use $\Upsilon_{\star \rm DYN}$ values for the (`$\star$ + CEN') mass models from \citet{Loubser2020a}, i.e.\  where a dark matter mass component is not included in the dynamical modelling. We show this in Appendix \ref{inner5kpc} and find that it does not change our conclusions. 

As an additional test, we estimate a parametrised $\Upsilon_{\star \rm DYN}$ (to vary as a function of radius) following the results for M87 from \citet{Sarzi2018} (in Appendix \ref{varML}). We estimate the r-band $\Upsilon_{\star \rm DYN}$ ratio at 2.5 kpc (for the inner aperture 0 to 5 kpc) to be 50 per cent higher than at 10 kpc (outer aperture of 5 to 15 kpc). We re-run our dynamical modelling from \citet{Loubser2020a}, and show our findings in Figure \ref{varMLFig}, and find that it does not change our conclusions. Our results strongly suggest that there is substantial scatter in the IMF among the most massive early-type galaxies. It is unlikely that any other systematic over- or underestimation of $\Upsilon_{\star \rm DYN}$ from assumptions in our dynamical models (see e.g. the studies of \citealt{Thomas2007a, Thomas2007b, Li2016}) will change this result.

\subsubsection{$\Upsilon_{\star \rm POP}$}

To assess possible systematic errors in our stellar population analysis, and the possible effect on $\Upsilon_{\star \rm POP}$, we also use the full spectral fitting code FIREFLY (Fitting IteRativEly For Likelihood analYsis) as described in \citet{Wilkinson2017} and applied to 2 million SDSS DR14 and DEEP DR4 spectra in \citet{Comparat2017} and to MaNGA data in \citet{Goddard2017}. 

FIREFLY is a minimization fitting code that fits combinations of single-burst stellar population models by using an iterative best-fitting process and Bayesian methods. No priors are applied, and all solutions (and their weight) are retained within a statistical cut. Moreover, no multiplicative or additive polynomials are used to adjust the spectral shape (as used in ULySS), and the continuum information is retained and used to determine the parameters. FIREFLY is compared to STARLIGHT \citep{CidFernandes2005}, STECKMAP \citep{Ocvirk2006} and VESPA \citep{Tojeiro2009} in \citet{Wilkinson2017}.

We use the MaStar stellar population models \citep{Maraston2020}, built from the MaNGA stellar library \citep{Yan2019}, with the empirical (E-MaStar) stellar library (see \citealt{Maraston2020}) due to its coverage in age and metallicity parameter space and high resolution. We derive light-weighted SSP-equivalent ages and metallicities for a Salpeter and a Kroupa IMF, similar to our method in ULySS (but with no priors on age components), and we use the stellar population results to derive the $\Upsilon_{\star \rm POP}$ in the $r$-band. 

In Appendix \ref{MaStar} (Figure \ref{ML_ML_FIREFLY}), we illustrate and describe how Figure \ref{ML_ML} changes using a different stellar population model, stellar library, and full spectrum fitting method. The comparison in Appendix \ref{MaStar} indicates that realistic errors on $\Upsilon_{\star \rm POP}$ should be larger to include the systematic errors from using a different combination of stellar population model, library, and fitting method. Even though using a different stellar population analysis has a pronounced effect on the determination of $\Upsilon_{\star \rm POP}$, it does not eliminate the variety of IMFs necessary to describe the BCGs. In Figure \ref{ML_ML_AVERAGE}, we show that the average (and standard deviation) of the two different determinations of $\Upsilon_{\star \rm POP}$ still scatter above and below the 1-to-1 line. This standard deviation is used as error bars in Figure \ref{ML_ML} to indicate realistic systematic uncertainties on $\Upsilon_{\star \rm POP}$. We further briefly compare the ages and metallicities derived using six different combinations of stellar population model, library, and fitting method in Appendix \ref{MaStar}, and find that no single combination can consistently derive stellar population parameters that would reconcile the $\Upsilon_{\star \rm POP}$ above as well as below the 1-to-1 line with $\Upsilon_{\star \rm DYN}$.

\subsection{Galaxies with low $\Upsilon_{\star \rm DYN}$}
\label{low_dyn}

We now consider the four BCGs below the 1-to-1 comparison line in Figure \ref{ML_ML} individually. The BCG in Abell 1689 has $\Upsilon_{\star \rm DYN}$ of $3.60 \pm 0.09$, which is not notably lower than what we expect for passively evolving BCGs, and Figure \ref{ML_ML_AVERAGE} shows that if we take the average of two stellar population method/models (ULySS/Vazdekis/MILES and FIREFLY/MaStar/E-MaStar) then the error bars on $\Upsilon_{\star \rm POP}$ are larger and both the Salpeter and Kroupa IMF reach the 1-to-1 line reconciling the dynamical and stellar population estimates. For Abell 754 ($\Upsilon_{\star \rm DYN} = 2.56$), with one of the lowest dynamical mass estimates, and a corresponding low central velocity dispersion (295 $\pm$ 14 km s$^{-1}$), Figure \ref{ML_ML_AVERAGE} also shows that the average of two stellar population methods/models reconciles the dynamical and stellar population estimates.

For two BCGs the dynamical and stellar population estimates of $\Upsilon_{\star}$ could not be reconciled in any way. Abell 1942 ($\Upsilon_{\star \rm DYN} = 1.23$) has an SSP-equivalent age of $\sim$4 Gyr, and therefore had more recent star formation (but not enough or recent enough to constrain the young stellar component, see \citealt{Loubser2016}). Abell 1942 is also bright (M$_{K}$ = --27.40 mag, compared to the average in our sample M$_{K} = -26.52$ mag), but with a below-average central velocity dispersion (296 km s$^{-1}$) compared to the other BCGs. For the BCG in Abell 1763 ($\Upsilon_{\star \rm DYN} = 1.89$), we also find that it has one of the lowest contributions of dark matter mass to total mass in the centre (3.4 per cent) on account of its brightness (M$_{K}$ = --27.33 mag) and high stellar mass. 

Therefore, the BCGs of Abell 1942 and 1763 are peculiar in that they are very bright (i.e. high stellar mass, and low $\Upsilon_{\star \rm DYN}$) compared to other BCGs of similar velocity dispersion. This is similar to the findings of \citet{Newman2017} (their figure 2) for the SNELLS systems, which have a total $\Upsilon$ lower than that for galaxies with similar mass and fall below the expectation from $\Upsilon$ versus velocity dispersion scaling relation (i.e.\ a projection of the Fundamental Plane). Their results show that the SNELLS galaxies also have peculiar properties that are not related to possible issues with, e.g., the IMF and the stellar population analysis.

\medskip

As mentioned in Section \ref{introduction}, \citet{Smith2014} concluded that the IMF constraints for individual galaxies, determined using different techniques ($\Upsilon_{\star \rm POP}$ vs $\Upsilon_{\star \rm DYN}$), do not always correlate on a galaxy-by-galaxy basis. However, \citet{Lyubenova2016} used the CALIFA sample to show that stellar populations and stellar dynamics give consistent results for a systematically varying IMF, and their results strongly suggest no case-to-case inconsistencies. They emphasise that inconsistencies found in other studies may be due to differences in stellar population models or aperture sizes used, or non-optimal dark matter halo corrections. In this paper, our direct comparison between $\Upsilon_{\star \rm POP}$ and $\Upsilon_{\star \rm DYN}$, and the search for possible correlations with other properties (Section \ref{corr}), assumes a consistency between the dynamical and stellar population determined $\Upsilon_{\star}$, and no case-by-case variations.


\section{Correlations of the mass-excess factor with other properties}
\label{corr}

When comparing mass-to-light ratios from dynamical mass models to those from stellar populations, the constraint on the IMF is often expressed as the mass-excess factor $\alpha =  \log_{10} ( \Upsilon_{\star \rm DYN}) -  \log_{10} ( \Upsilon_{\star \rm POP})$,  where $\Upsilon_{\star \rm POP}$ is most-commonly the predicted stellar mass-to-light ratio for some reference IMF, in our case for a Salpeter IMF \citep{Treu2010}. We present the mass-excess factor and the propagated uncertainty in Table \ref{Alpha} (column 2 and 3) for the fourteen BCGs from Figure \ref{ML_ML}. If we assume no case-to-case inconsistencies as described above, a mass-excess factor of $\alpha=0$ implies the galaxy has a stellar mass-to-light ratio in agreement with the chosen reference IMF. An $\alpha>0$ indicates departures from this IMF which could be either a bottom- or top-heavy IMF due to a higher fraction of low mass stars or stellar remnants, respectively. As mentioned in Section \ref{Constraints_IMF}, for passively-evolving, massive early-type galaxies consisting of old, low-mass stars, it points to a bottom-heavy IMF.

\begin{table}
\caption{The mass-excess factor $\alpha =  \log_{10} ( \Upsilon_{\star \rm DYN}) -  \log_{10} ( \Upsilon_{\star \rm POP})$ and the error, where $\Upsilon_{\star \rm POP}$ is the predicted stellar mass-to-light ratio using a Salpeter IMF (column 2 and 3), and $\Upsilon_{\star \rm DYN}$ is the stellar mass-to-light ratio from dynamics where the dark matter mass component was included (<15 kpc). For reference we also include $\alpha$ where $\Upsilon_{\star \rm POP}$ is derived using a Kroupa IMF (column 4 and 5).}
\label{Alpha}
\centering
\begin{tabular}{lrrrr}
\toprule
BCG  & $\alpha_{\rm S, DM}$ & $\delta \alpha_{\rm S, DM}$ & $\alpha_{\rm K, DM}$ & $\delta \alpha_{\rm K, DM}$ \\
\hline
A68   & 0.145 & 0.035 & 0.348 & 0.032  \\
A267 & 0.104 & 0.014 & 0.270 & 0.022  \\ 
A646  &  0.260 & 0.071 & 0.427 & 0.038 \\
A754  &  --0.239 & 0.026 & --0.083 & 0.023 \\
A1650  & 0.191 & 0.029 & 0.391 & 0.020  \\	
A1689  & --0.137 & 0.018 & 0.026 & 0.029  \\	
A1763  & --0.366 & 0.024 & --0.140 & 0.036  \\
A1942  & --0.366 & 0.025 & --0.177 & 0.031 \\
A1991  & 0.016 & 0.085 & 0.179 & 0.075 \\
A2029 &  0.208 & 0.111 & 0.404 & 0.127 \\
A2050  & 0.113 & 0.058 & 0.334 & 0.021 \\
A2259  & 0.135 & 0.025 & 0.333 & 0.034 \\
A2261  & 0.062 & 0.036 & 0.242 & 0.040 \\
A2420  & 0.120 & 0.041 & 0.269 & 0.028  \\
\bottomrule
\end{tabular}
\end{table}

\begin{figure*}
\centering
   \subfloat{\includegraphics[scale=0.45]{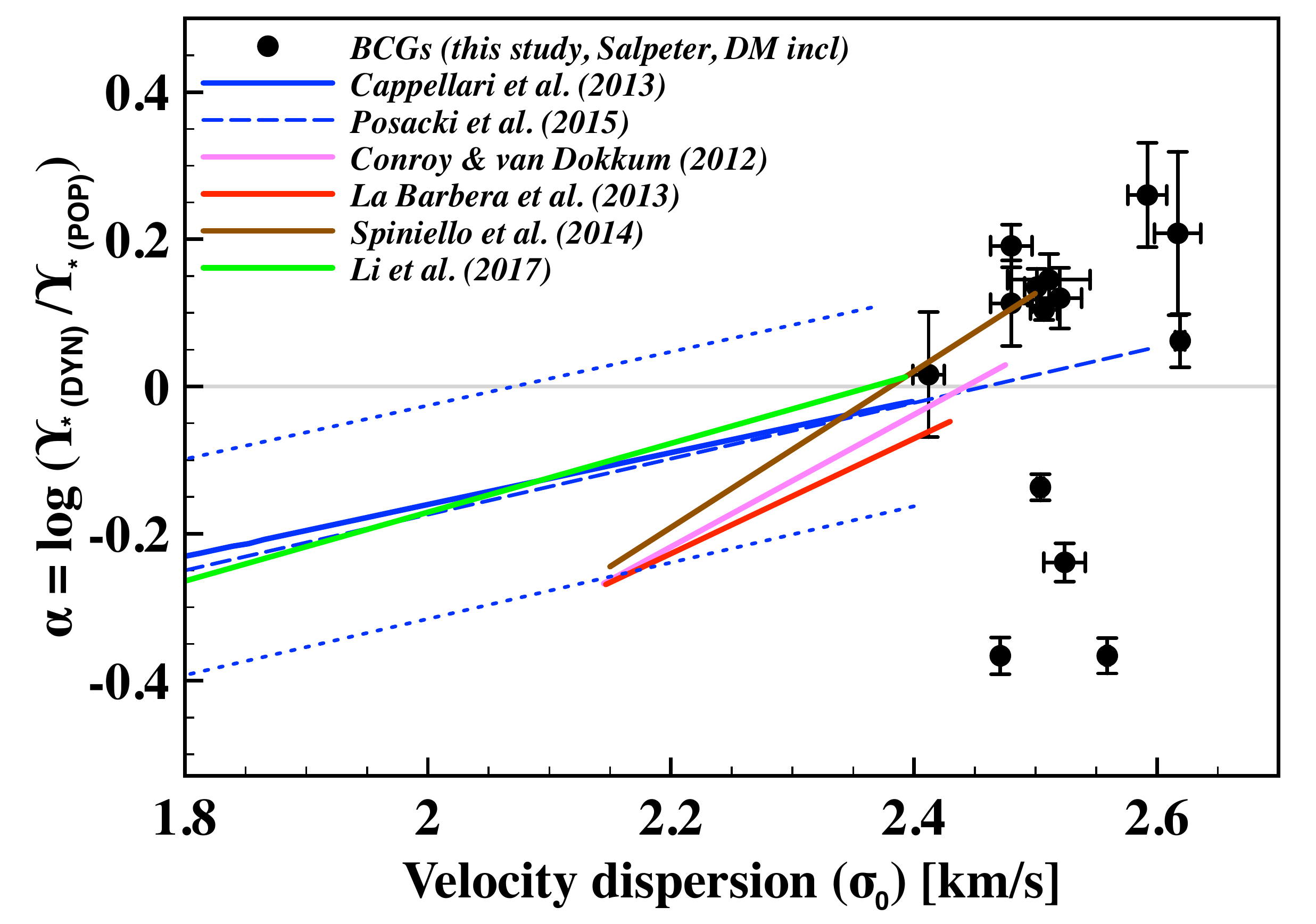}}
   \caption{We plot the mass-excess factor ($\alpha$) against velocity dispersion ($\sigma_{0}$), and indicate the BCGs (from Figure \ref{ML_ML}) with black symbols. The horizontal line at $\alpha = 0$ show the mass excess expected for a Salpeter IMF \citep{Barber2018}. The observed trend (solid line), as well as intrinsic scatter (dotted lines), from \citet{Cappellari2013} is shown in blue (measured from dynamical modelling). The dashed blue line shows the extension to the \citet{Cappellari2013} sample by \citet{Posacki2015} to include SLACS galaxies. The correlation by \citet{Li2017} (green line) is also measured from dynamical modelling, whereas the other correlations are measured from stellar population measurements \citep{Spiniello2014, Conroy2012, LaBarbera2013} (brown, pink and red lines).} 
\label{ML_MLE}
\end{figure*}

Although it is known that the mass-excess factor ($\alpha$) increases with velocity dispersion ($\sigma$) in early-type galaxies \citep{Cappellari2013, LaBarbera2015}, consensus has not yet been reached on if, and how, it varies with other galaxy properties \citep{Conroy2012, McDermid2014, LaBarbera2015, MartinNavarro2015a,MartinNavarro2015b}. We plot the $\alpha$ of the fourteen BCGs from Table \ref{Alpha} against velocity dispersion in Figure \ref{ML_MLE}. As the fourteen BCGs are all very massive, our range in velocity dispersion is too narrow to measure a correlation with $\alpha$. 

However, we compare our data points to correlations from the literature for early-type galaxies, derived using different methods, following the compilation summarised in \citet{Barber2018} (their figure 1). In Figure \ref{ML_MLE} we include, in blue, the observed relation (for $r$-band) between mass excess (also for a Salpeter reference IMF) with velocity dispersion measured in \citet{Cappellari2013} for massive elliptical galaxies in the ATLAS3D survey. Also shown are observed trends from \citet{Conroy2012}, \citet{LaBarbera2013} and \citet{Spiniello2014}. We also show fits from \citet{Li2017} for elliptical and lenticular galaxies using two different stellar population models and a Salpeter IMF. The dashed blue line shows the extension to the ATLAS3D \citep{Cappellari2013} sample by including SLACS galaxies \citep{Posacki2015}. This mass-excess -- velocity dispersion relation from \citet{Posacki2015} is slightly steeper than from the ATLAS3D sample alone, suggesting a non-linear relation that depends on the range of velocity dispersion probed.

From Figure \ref{ML_MLE}, we see that the variations in $\alpha$ are larger than any measurement errors and strongly suggests inconsistency with a single universal IMF. We note that our BCGs are generally more massive than the velocity dispersion range covered by previous studies of massive ellipticals. We emphasise that we excluded the BCGs where we see evidence for a radially variant $\Upsilon_{\star}$ in their stellar population properties. We also note that our quantities ($\sigma_{0}$, $\alpha$) were measured in the central part of the BCGs (which is less affected by the dark matter contribution), whereas the quantities from the reference correlations were measured within the effective radius ($R_{e}$). Nevertheless, dark matter mass, as constrained from weak lensing observations, are included in our dynamical models (see \citealt{Loubser2020a}). The velocity dispersions measured for the reference correlations ($\sigma_{e}$ within $R_{e}$) trace the overall galaxy potential (mass) instead of the detailed kinematics. For the BCGs, we found that the velocity dispersion gradients are very diverse \citep{Loubser2018}, and can be steep within their large effective radii \citep{Loubser2020a}. However, we use this information and do an aperture correction to derive the velocity dispersion within the half-light radius, $\sigma_{e}$, for the BCGs. We show this plot in Appendix \ref{apertures}, and find that it does not influence our conclusions. For the BCGs better described by a Salpeter (or heavier) IMF, our data points fall on an extrapolation of the correlations, also suggesting a systematic variation of the IMF for these galaxies in that an increasingly bottom-heavy IMF is needed for more massive galaxies (higher $\sigma$), as opposed to case-by-case inconsistencies between the dynamical and stellar population methods. This figure also emphasises the substantial scatter in the IMF among the most massive galaxies.

\begin{figure*}
\centering
   \subfloat{\includegraphics[scale=0.22]{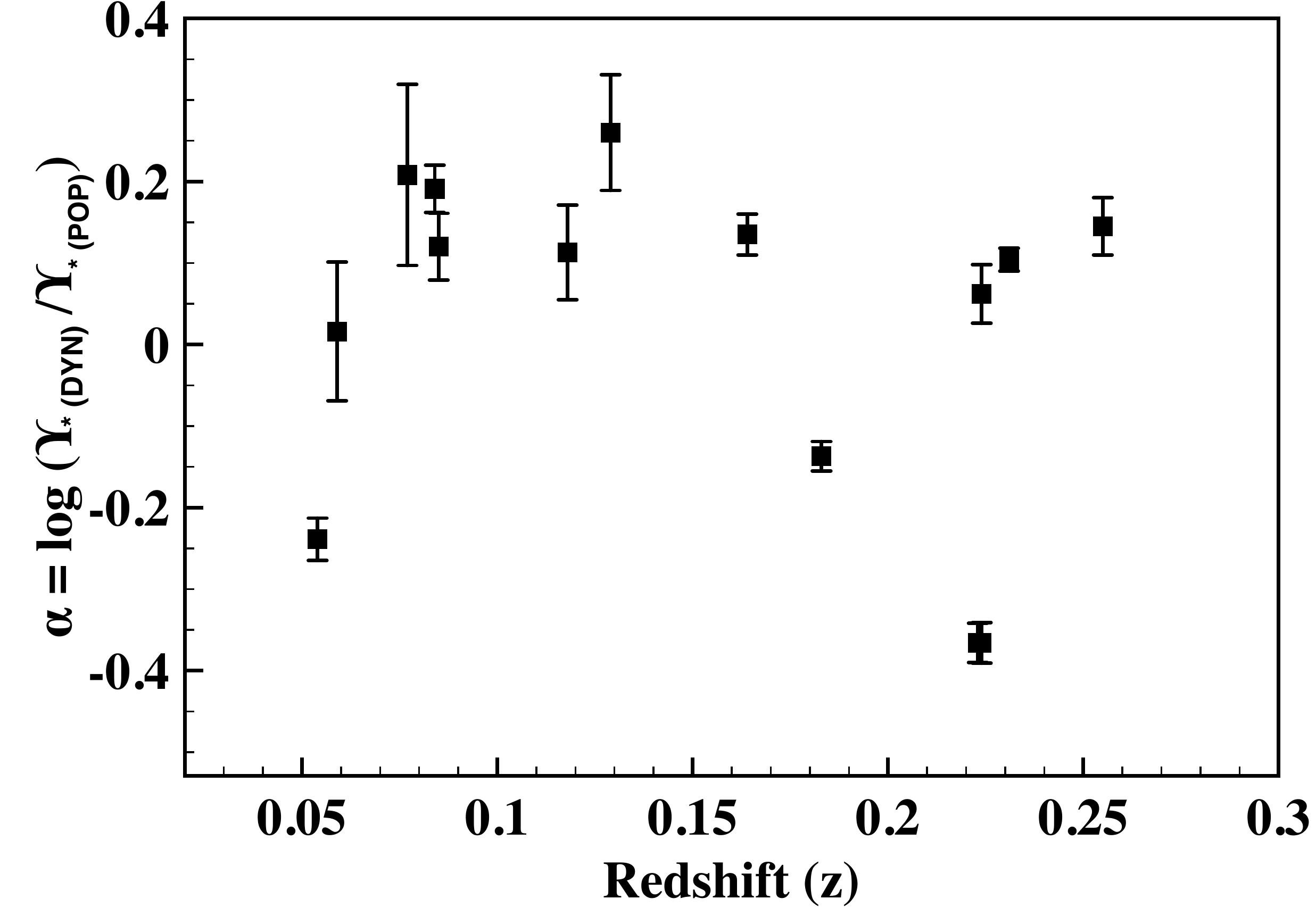}}
   \subfloat{\includegraphics[scale=0.22]{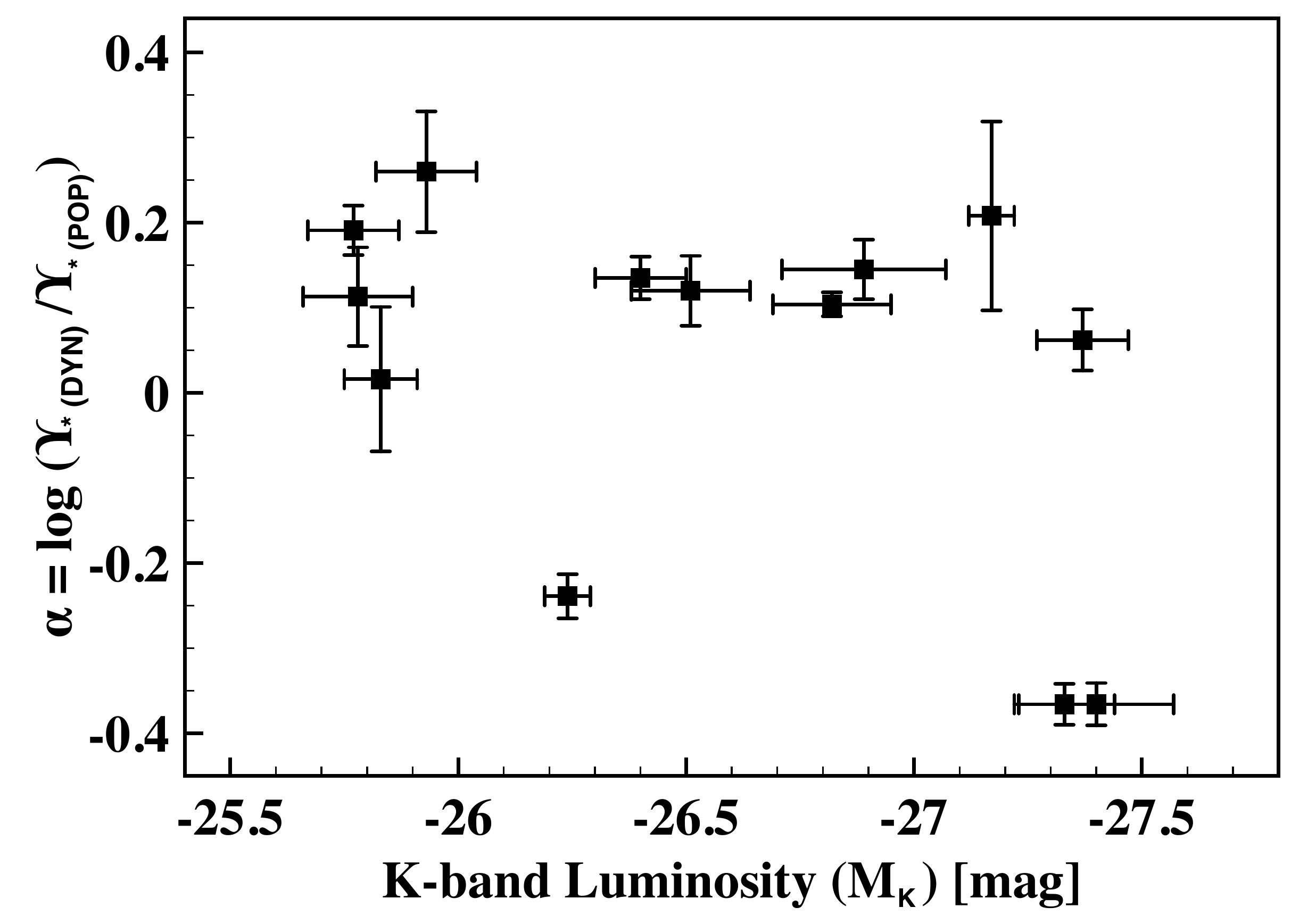}}
   \subfloat{\includegraphics[scale=0.22]{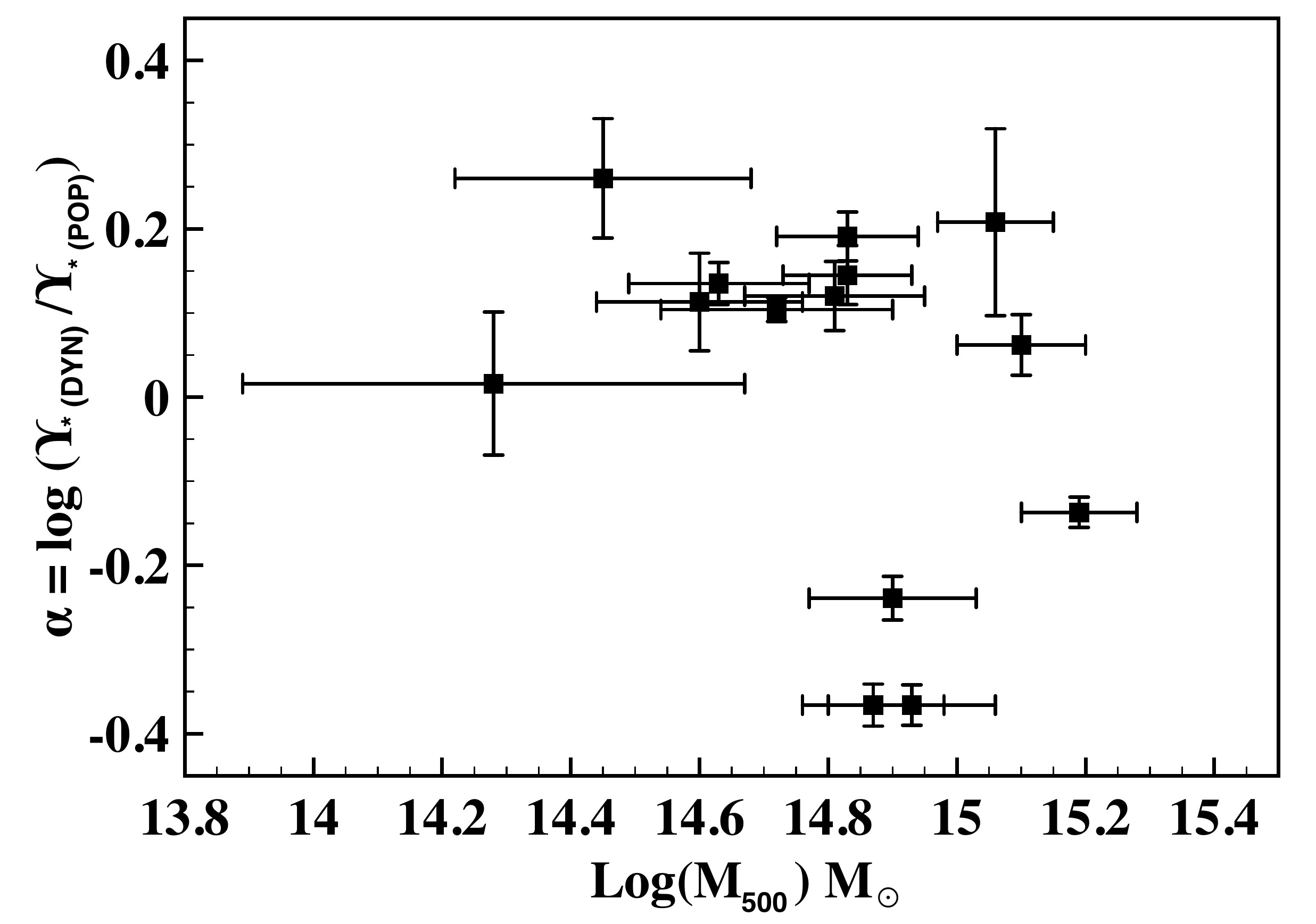}}\\
   \caption{We plot the mass-excess factor ($\alpha$) against redshift, $K$-band luminosity (if available, see \citet{Loubser2016}), and $M_{500}$ \citep{Herbonnet2019} for the BCGs. We see no clear correlation between the mass-excess factor and any of these properties.}
\label{Correlations}
\end{figure*}

For the fourteen BCGs from Section \ref{15}, we further show (Figure \ref{Correlations}) the mass excess factor ($\alpha$) against redshift, $K$-band luminosity, and $M_{500}$, and find no correlations. Some previous studies e.g. \citet{McDermid2014, Davis2017} do not find any correlations between their IMF mass-excess parameter and galaxy dynamical or stellar population properties. Other studies find a correlation with metallicity, e.g. \citet{MartinNavarro2015a, MartinNavarro2015b, Zhou2019}, or find a correlation with [$\alpha$/Fe] abundances, e.g. \citet{Conroy2012}, both properties that are strongly correlated with the velocity dispersion. These correlations, or lack thereof, are better studied using large samples of galaxies over a wide mass range \citep{Posacki2015}, or cosmological simulations \citep{Sonnenfeld2017, Barber2018}.

In future, it will be interesting to use a physical parameterisation instead of the mass excess factor. An example is F$_{0.5}$, defined as the fraction of stars with masses below 0.5 M$_{\sun}$, often used for massive early-type galaxies \citep{LaBarbera2013, MartinNavarro2015b, LaBarbera2015, Lyubenova2016, MartinNavarro2019}. However, the interpretation of this is vastly complicated by the fact that spectroscopic IMF studies are sensitive to the present-day stellar populations, and the large variety of parametrisations of IMF variations makes comparison between different methods difficult. Cosmological, hydrodynamical simulations (e.g.\ \citealt{Clauwens2016, Barber2018}) are needed to aid our interpretation of physical parameterisations derived from observed properties. 


\section{Conclusions}
\label{conclusions}

We investigate the stellar and dynamical mass profiles as well as stellar populations of BCGs, and use the results to place constraints on the stellar IMF. In an accompanying paper \citep{Loubser2020a}, we have modelled the stellar and dynamical masses of 25 BCGs using the MGE \citep{Emsellem1994, Cappellari2002} and JAM \citep{Cappellari2006, Cappellari2008} methods, deriving the stellar mass-to-light ratio ($\Upsilon_{\star \rm DYN}$), and stellar velocity anisotropy ($\beta_{z}$), where the dark matter mass was constrained from weak lensing results. Here, we study the spatially-resolved stellar population properties of the BCGs, and use it to calculate their stellar mass-to-light ratios ($\Upsilon_{\star \rm POP}$) assuming a (single power law) Salpeter and (double power law) Kroupa IMF. We compare the stellar mass-to-light ratios derived from the two independent methods ($\Upsilon_{\star \rm POP}$ with $\Upsilon_{\star \rm DYN}$) and use it to constrain the IMF. We summarise our main conclusions as follows:

\begin{enumerate}
\item Of the fourteen MENeaCS BCGs, we find three BCGs (Abell 780, Abell 1795 and Abell 2055) with very young stellar populations ($<$200 Myr) in their inner (0 -- 5 kpc) apertures. Together with the four of the eighteen CCCP BCGs for which we have detected young stellar populations in \citet{Loubser2016}, it constitutes 22 per cent of the full sample of 32 BCGs, equally distributed in redshift. From Figure \ref{fig:ML} it follows, similar to our conclusions in \citet{Loubser2016}, that the BCGs with young stellar populations are located in host clusters with short cooling times.

\item We use the ages and metallicities derived from our spectra to determine $\Upsilon_{\star \rm POP}$ in the $r$-filter in an inner (0 -- 5 kpc) and outer (5 -- 15 kpc) aperture, and find that 19/32 (60 per cent) of the BCGs have constant $\Upsilon_{\star \rm POP}$ over this radial range (0 -- 15 kpc). The non-constant $\Upsilon_{\star \rm POP}$ in these BCGs are primarily driven by age gradients between the inner and outer apertures (i.e.\ inner aperture significantly younger than the outer aperture). 

\item To place constraints on the IMF, we eliminate: i) the two BCGs for which we find extreme tangential anisotropy ($\beta_{z} < -1$, see \citealt{Loubser2020a}); ii) the BCGs where we detect young stellar components; iii) the BCGs where we detect significant age gradients between the inner and the outer stellar population bin (i.e. $\Upsilon_{\star \rm POP}$ is non-constant within 15 kpc). We compare $\Upsilon_{\star \rm POP}$ vs $\Upsilon_{\star \rm DYN}$ (with the inclusion of a dark matter mass component) for 14 BCGs in Figure \ref{ML_ML}.  

From this comparison (Figure \ref{ML_ML}), it follows that the majority of the BCGs are better described by a bottom-heavy Salpeter or heavier IMF, but there is a small number of data points below the 1-to-1 line, for which a Kroupa IMF describes the data much better. This agrees with the studies for massive early-type galaxies that find a `bottom-heavy' (Salpeter-like) IMF \citep{Cappellari2013, LaBarbera2015, MartinNavarro2015a} as well as with the SNELLS galaxies which instead measured $\Upsilon_{\star}$ consistent with a Milky-Way (Kroupa-like) IMF  \citep{Smith2017, Newman2017, Collier2018}, and confirms substantial scatter in the IMF among the most massive galaxies. 

We test various possible systematic effects on this direct comparison (e.g. only considering the inner aperture, or using a variable $\Upsilon_{\star \rm DYN}$ in our dynamical models) in the Appendices. Although using different stellar population analysis models and a different fitting method can bring some data points in better agreement with the 1-to-1 line, none of these systematic effects can consistently reconcile all the $\Upsilon_{\star \rm POP}$ and $\Upsilon_{\star \rm DYN}$ measurements above and below the 1-to-1 line.

\item We plot the mass-excess factor against velocity dispersion in Figure \ref{ML_MLE}, and compare it to correlations from the literature for massive elliptical galaxies, derived using different methods (also see \citealt{Barber2018}). For the BCGs better described by a Salpeter (or heavier) IMF, our data points fall on an extrapolation of the correlations, also suggesting a systematic variation of the IMF for these galaxies (as opposed to a case-by-case inconsistencies).
\end{enumerate}

In summary, we find substantial scatter in the IMF among the most massive galaxies. For most BCGs, a Salpeter (or even more bottom-heavy) IMF is required. For one BCG a Kroupa IMF is preferred, and for another two BCGs an IMF even lighter than a Kroupa IMF is preferred. Our dark matter fractions are consistent with previous studies (on average), and even though including more mass attributed to dark matter will bring some of the BCGs above the 1-to-1 line closer to the line, it will move the four BCGs better described by a Kroupa IMF further below the 1-to-1 comparison line. A universal IMF will therefore not only be inconsistent with the weak lensing mass measurements, but will also imply very high dark matter fractions for some BCGs (within our limited radial range of $<$15 kpc) and none for others. 

\section*{Acknowledgements}

We gratefully acknowledge the constructive report from the anonymous reviewer. This research was enabled, in part, by support provided by the bilateral funding agreement between the National Research Foundation (NRF) of South Africa, and the Netherlands Organisation for Scientific Research (NWO) to SIL and HH. AB acknowledges support from NSERC (Canada) through the Discovery Grant program. HH acknowledges support from NWO through VICI grant 639.043.512. YMB acknowledges funding from the EU Horizon 2020 research and innovation programme under Marie Sk{\l}odowska-Curie grant agreement 747645 (ClusterGal) and the NWO through VENI grant 639.041.751. 

Based, in part, on observations obtained at the Gemini Observatory, which is operated by the Association of Universities for Research in Astronomy, Inc., under a cooperative agreement with the NSF on behalf of the Gemini partnership: the National Science Foundation (United States), the National Research Council (Canada), CONICYT (Chile), Ministerio de Ciencia, Tecnolog\'{i}a e Innovaci\'{o}n Productiva (Argentina), and Minist\'{e}rio da Ci\^{e}ncia, Tecnologia e Inova\c{c}\~{a}o (Brazil). 

Based, in part, on observations obtained at the Canada-France-Hawaii Telescope (CFHT) which is operated by the National Research Council of Canada, the Institut National des Sciences de l'Univers of the Centre National de la Recherche Scientifique of France, and the University of Hawaii. This research used the facilities of the Canadian Astronomy Data Centre operated by the National Research Council of Canada with support from the Canadian Space Agency. 

Any opinion, finding and conclusion or recommendation expressed in this material is that of the author(s) and the NRF does not accept any liability in this regard.

\section*{Data availability}

Based, in part, on observations obtained at the Gemini Observatory (GN-2008A-Q103, GN-2008B-Q5, GN-2009A-Q107, GN-2009B-Q118, GS-2008A-Q21, GS-2008B-Q4, GS-2009A-Q82). We use versions of the publicly available ULySS \citep{Koleva2009}, and FIREFLY \citep{Wilkinson2017} software packages, modified for our data and analysis. This research made use of Astropy\footnote{http://www.astropy.org}, a community-developed core Python package for Astronomy \citep{astropy:2013,astropy:2018}. The data underlying this article will be shared on reasonable request to the corresponding author.




\bibliographystyle{mnras}
\bibliography{References}



\appendix

\section{Cylindrically- or Spherically-aligned Jeans Axisymmetric Models}
\label{spherical}

In addition to the axisymmetric Jeans equations for cylindrically-aligned coordinates, we also use the axisymmetric Jeans equations for spherically-aligned coordinates \citep{Cappellari2020}, since a comparison between the two solutions allow for a robust assessment of the modelling results and dynamical parameters. We adapt the spherically-aligned JAM models (abbreviated as JAM$_{\rm sph}$) for our purpose by modifying the models to fit our data, and to include a dark matter mass component.

\begin{figure}
\centering
   \includegraphics[scale=0.42]{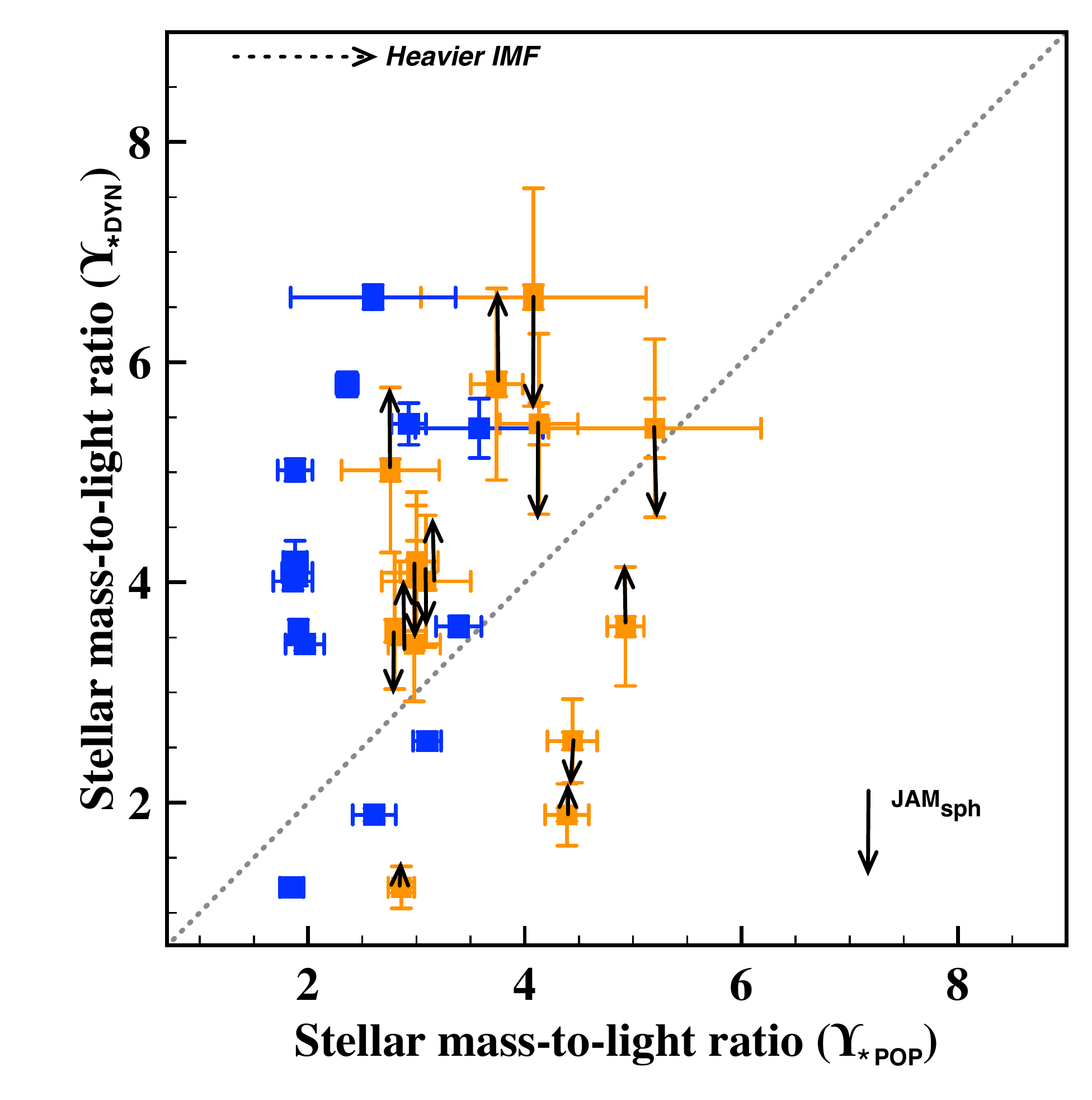}
   \caption{$\Upsilon_{\star \rm DYN}$ (dark matter included) vs $\Upsilon_{\star \rm POP}$ for Salpeter (orange) and Kroupa (blue) IMFs, as discussed in Section \ref{15}. The solid black arrows indicate a 15 per cent change in $\Upsilon_{\star \rm DYN}$ from using JAM$_{\rm sph}$ instead of JAM ($\Upsilon_{\star \rm DYN}$ is lower for decreasing $\nu_{\rm rms}$ profile BCGs, i.e. radial anisotropy, and $\Upsilon_{\star \rm DYN}$ is higher for increasing $\nu_{\rm rms}$ profile BCGs, i.e. tangential anisotropy). The dotted line indicates the 1-to-1 line, and the arrow in the top left corner indicate that a heavier IMF moves the data points to higher values on the x-axis.}
\label{ML_ML_sph}
\end{figure}

In \citet{Loubser2020a}, we found that neither model is significantly, or consistently, better or worse than the other. However, we did find that JAM$_{\rm sph}$ is relatively insensitive to anisotropy (i.e. a bigger change from $\beta = 0$ is required to best fit the observed kinematics). Corresponding to this systematic change in velocity anisotropy in JAM$_{\rm sph}$, there is a systematic change in best-fitting $\Upsilon_{\star \rm DYN}$, where $\Upsilon_{\star \rm DYN}$ is lower for decreasing $\nu_{\rm rms}$ profile BCGs (i.e. radial anisotropy, or $\beta$ positive), and higher for increasing $\nu_{\rm rms}$ profile BCGs (i.e. tangential anisotropy, or $\beta$ negative). These changes are larger than the statistical error on the parameters, and can correspond to a change of up to 15 per cent in best-fitting $\Upsilon_{\star \rm DYN}$. We indicate a 15 per cent uncertainty, and the direction of the change (whether it is higher or lower $\Upsilon_{\star \rm DYN}$) in Figure \ref{ML_ML_sph}, and find that it will move data points both above and below the 1-to-1 line higher and lower, and that this assumption we made in the dynamical modelling does not affect our conclusions.  

\section{The inner 5 kpc}
\label{inner5kpc}

\begin{figure}
\centering
   \includegraphics[scale=0.42]{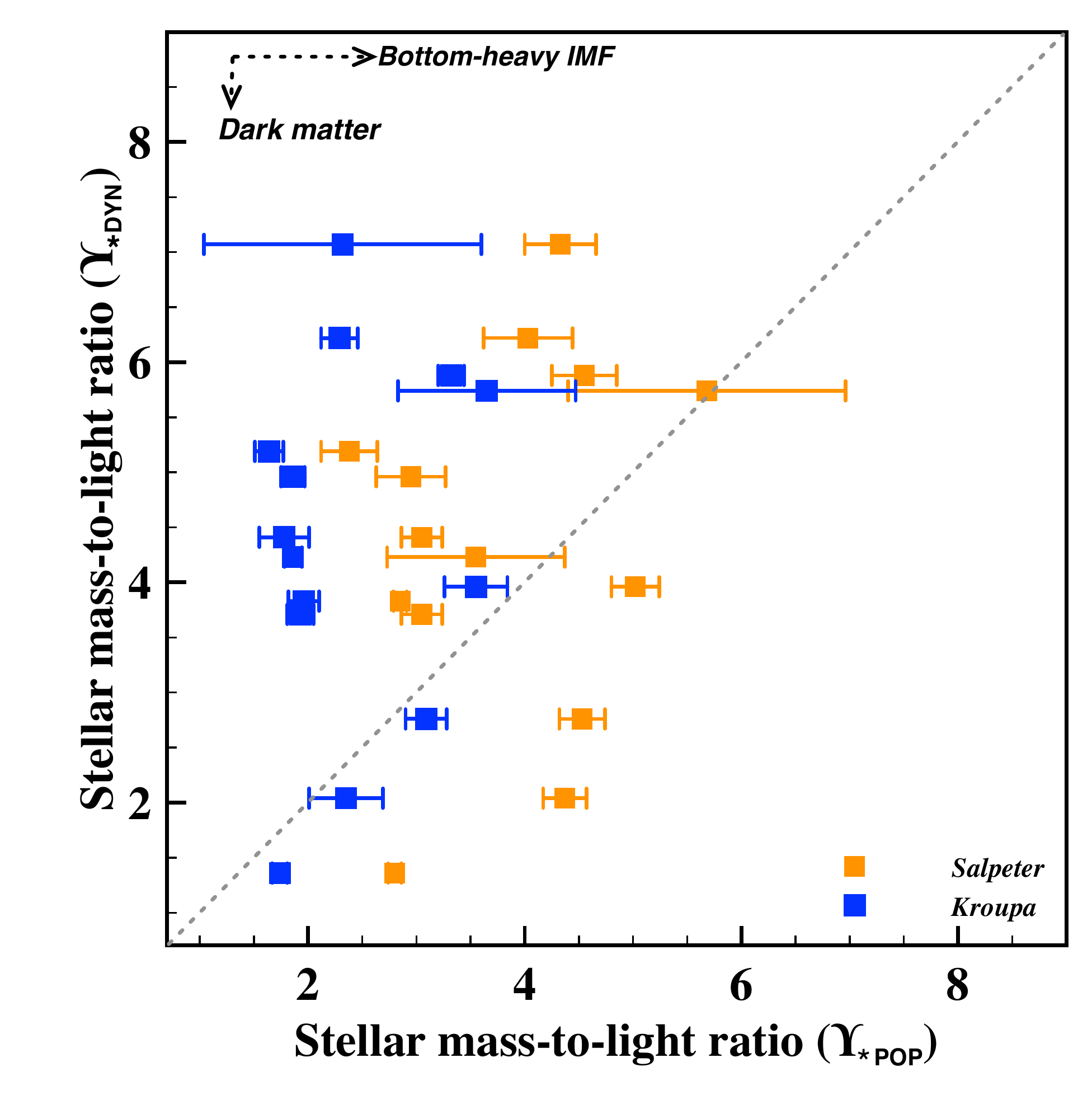}
   \caption{$\Upsilon_{\star \rm DYN}$ (dark matter not included) vs $\Upsilon_{\star \rm POP}$ (for just the inner aperture) for Salpeter (orange) and Kroupa (blue) IMFs for the inner 5 kpc, as discussed in Section \ref{systematic_errors}. The dotted line indicates the 1-to-1 line, and the arrows in the top left corner indicate that adding dark matter move the data points to lower values on the y-axis, and using a more bottom-heavy IMF moves the data points to higher values on the x-axis.}
\label{ML_ML_inner}
\end{figure} 

We also investigate the comparison between $\Upsilon_{\star \rm DYN}$ and $\Upsilon_{\star \rm POP}$ (for a Salpeter and Kroupa IMF), using $\Upsilon_{\star \rm POP}$ for just the central 5 kpc of the BCGs. We expect the dark matter mass component to contribute very little to the total mass, so we use $\Upsilon_{\star \rm DYN}$ values for the (`$\star$ + CEN') mass models from \citet{Loubser2020a}, i.e.\  where a dark matter mass component is not included in the dynamical modelling. We still find that the majority of the BCGs are better described by a `heavier' IMF, but that four BCGs below the 1-to-1 line are better described by a Kroupa IMF.

\section{Dynamical models with variable $\Upsilon_{\star \rm DYN}$}
\label{varML}

\begin{figure*}
\centering
   \subfloat{\includegraphics[scale=0.40]{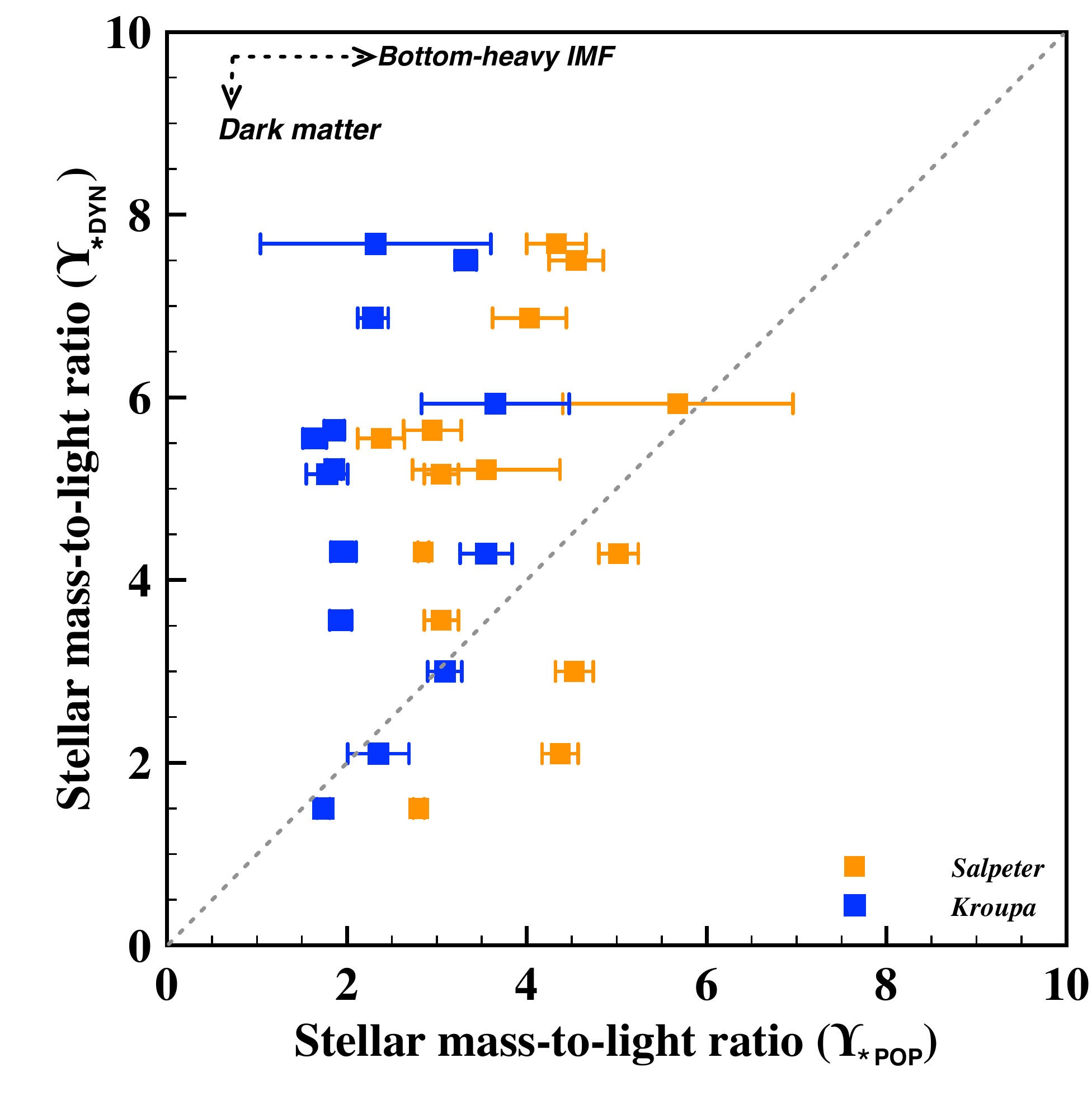}}
    \subfloat{\includegraphics[scale=0.40]{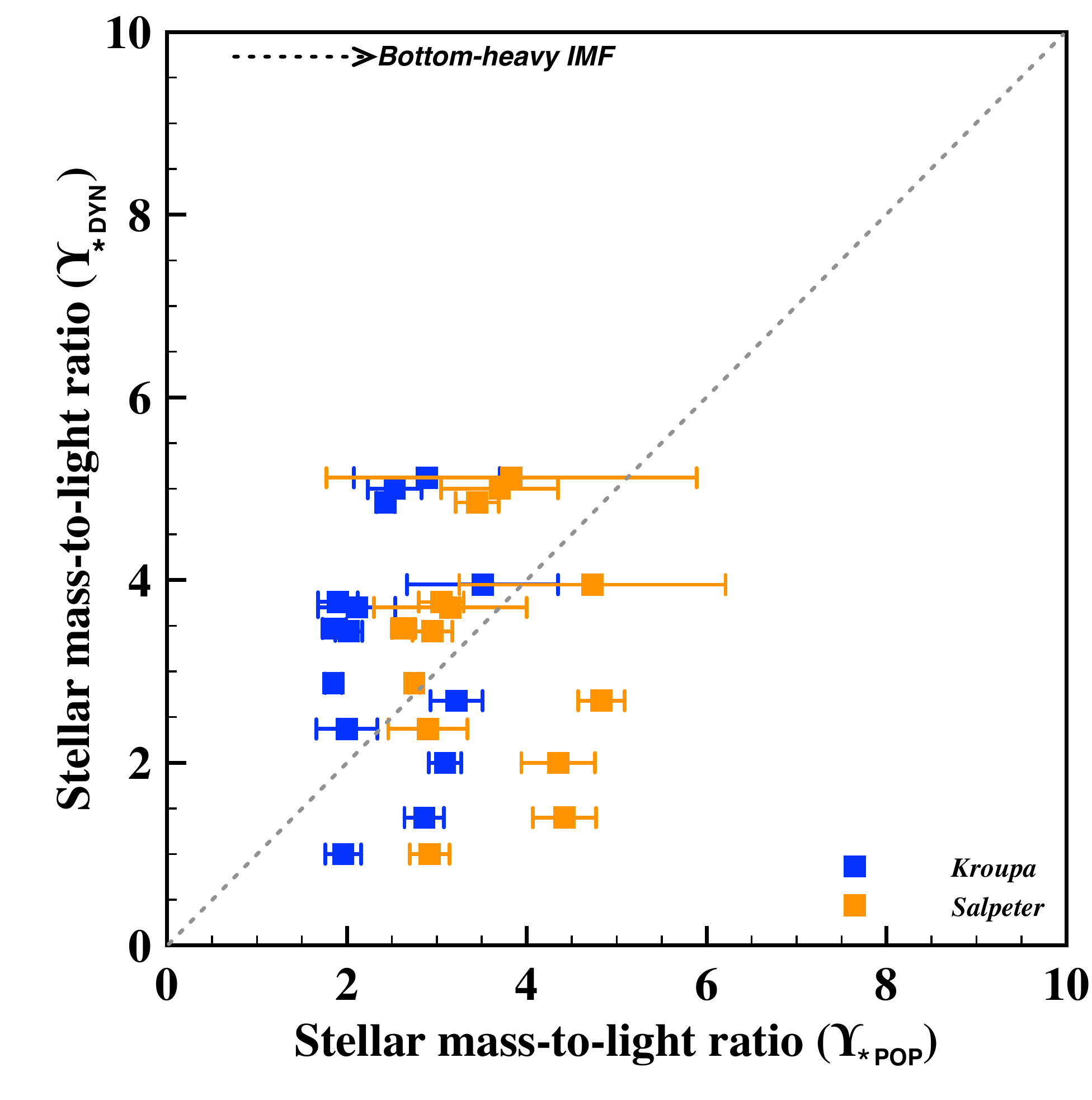}}
   \caption{Left: $\Upsilon_{\star \rm DYN}$ against $\Upsilon_{\star \rm POP}$ for the inner aperture (dark matter included, and additional dark matter will lower the values on the y-axis as indicated by the arrow. Right: $\Upsilon_{\star \rm DYN}$ against $\Upsilon_{\star \rm POP}$ for the outer aperture (dark matter included).}
\label{varMLFig}
\end{figure*}

As described in Section \ref{systematic_errors}, the IMF may also vary radially within high-mass early-type galaxies, becoming bottom-heavier towards the central regions \citep{VanDokkum2017, Oldham2018, Sarzi2018, Parikh2018, LaBarbera2019}. In addition to the test we do in Appendix \ref{inner5kpc}, we estimate a parametrised $\Upsilon_{\star \rm DYN}$ (to vary as a function of radius) following the results for M87 from  \citet{Sarzi2018}. Following their results (their Figure 11), we estimate the $r$-band $\Upsilon_{\star \rm DYN}$ ratio at 2.5 kpc (for the inner aperture 0 to 5 kpc) to be 50 per cent higher than at 10 kpc (outer aperture of 5 to 15 kpc). We note that \citet{Vaughan2018} find a constant IMF up to 0.7$R_{e}$ for the BCG NGC1399, which will imply a constant $\Upsilon_{\star \rm DYN}$ over our radial range. Nevertheless, we re-run our dynamical modelling from \citet{Loubser2020a}, and instead of adding additional free parameters, we find the best fitting $\Upsilon_{\star \rm DYN}$ (along with the best-fitting $\beta_{z}$) if we assume $\Upsilon_{\star \rm DYN}$ is 50 per cent larger in the inner aperture than in the outer aperture. We show our findings in Figure \ref{varMLFig}.

We find that the new best-fitting inner $\Upsilon_{\star \rm DYN}$ is higher than the constant overall (constant) best-fitting $\Upsilon_{\star \rm DYN}$ previously, and the outer $\Upsilon_{\star \rm DYN}$ is lower than the constant overall best-fitting $\Upsilon_{\star \rm DYN}$ previously, and that in general the new best-fitting $\beta_{z}$ is slightly lower than previously but it does not influence any conclusions from \citet{Loubser2020a}. A parametrised, variable $\Upsilon_{\star \rm DYN}$ does not systematically reconcile $\Upsilon_{\star \rm DYN}$ with $\Upsilon_{\star \rm POP}$ for all the BCGs, in the inner or the outer aperture, and for the Salpeter or the Kroupa IMF. 

\section{$\Upsilon_{\star \rm POP}$ derived from MaStar stellar population models using FIREFLY}
\label{MaStar}

\begin{figure}
\centering
   \subfloat{\includegraphics[scale=0.42]{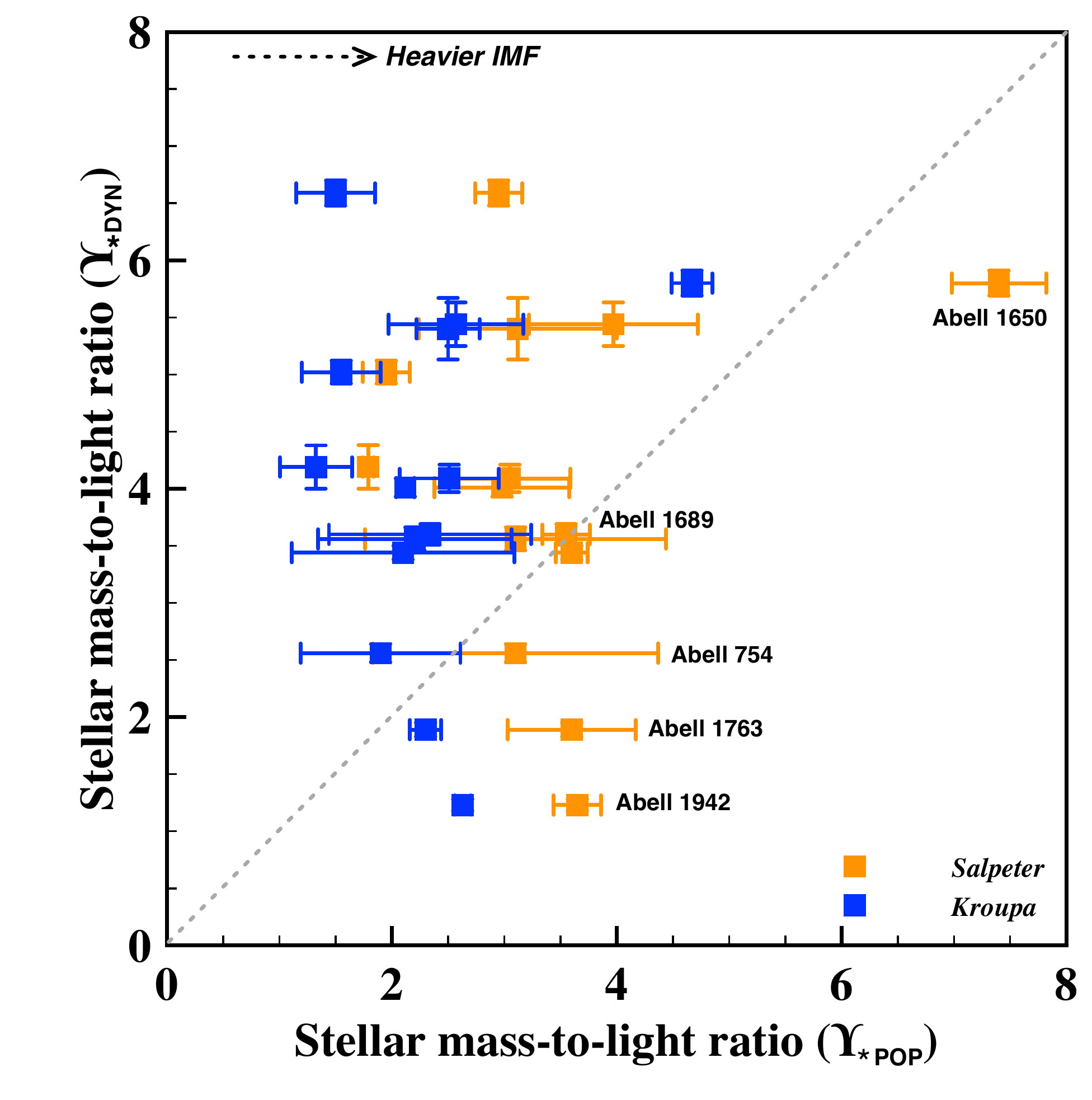}}
   \caption{$\Upsilon_{\star \rm DYN}$ (dark matter included) vs $\Upsilon_{\star \rm POP}$ for Salpeter (orange) and Kroupa (blue) IMFs, where $\Upsilon_{\star \rm POP}$ is from FIREFLY/MaStar, as discussed in Section \ref{systematic_errors}. } 
\label{ML_ML_FIREFLY}
\end{figure}

As described in Section \ref{systematic_errors}, we illustrate how Figure \ref{ML_ML} changes using a different stellar population model, stellar library, and full spectrum fitting method. We use FIREFLY (Fitting IteRativEly For Likelihood analYsis) as described in \citet{Wilkinson2017}, and the MaStar stellar population models with the empirical (E-MaStar) stellar library \citep{Maraston2020}. We derive light-weighted SSP-equivalent ages and metallicities for a Salpeter and a Kroupa IMF, similar to our method in ULySS (but with no priors on age components), and we use the stellar population results to derive the $\Upsilon_{\star \rm POP}$ in the $r$-band (Figure \ref{ML_ML_FIREFLY}).

\begin{figure}
\centering
   \subfloat{\includegraphics[scale=0.34]{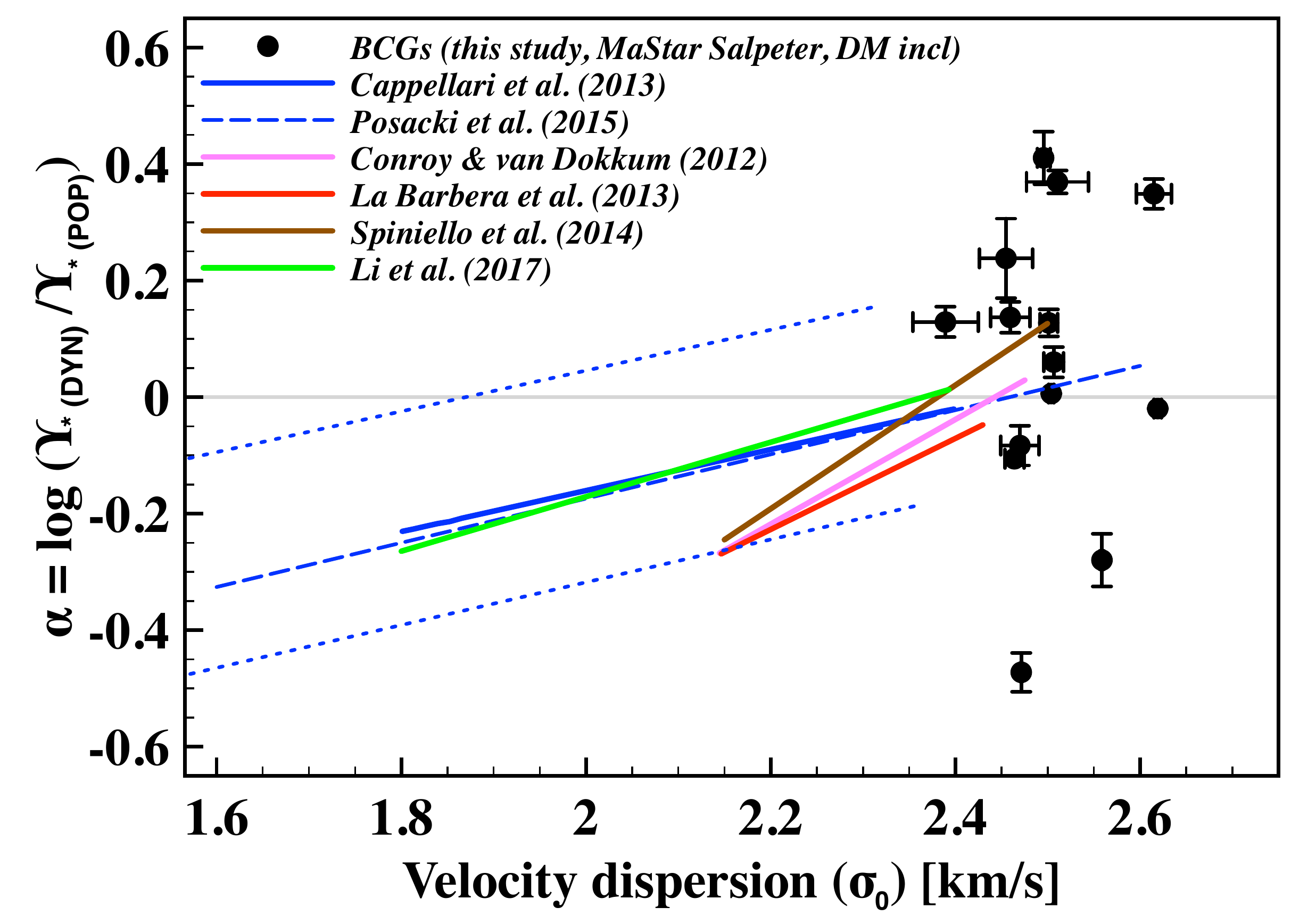}}
   \caption{We plot the mass-excess factor ($\alpha$, from FIREFLY/MaStar using a Salpeter IMF) against velocity dispersion $\sigma_{0}$, and indicate the BCGs with black symbols. The legend is the same as in Figure \ref{ML_MLE}.} 
\label{ML_FF}
\end{figure}

Of the four BCGs below the 1-to-1 line (Abell 754, 1689, 1763 and 1942), one BCG (Abell 1689) is now on the 1-to-1 line for a Salpeter IMF, and another BCG (Abell 754) now has error bars that also encompass the 1-to-1 line. For the two other BCGs (Abell 1763 and 1942), the Kroupa IMF is still more comparable to the 1-to-1 line, and in addition one other BCG (Abell 1650) is now also below the 1-to-1 line (for a Salpeter IMF). Of the 14 BCGs (using a Salpeter IMF), seven $\Upsilon_{\star \rm POP}$ determined using ULySS/Vazdekis/MILES and FIREFLY/MaStar/E-MaStar agree within the errors. Of the other seven, four $\Upsilon_{\star \rm POP}$ are smaller using FIREFLY/MaStar/E-MaStar, and three larger. We also illustrate how Figure \ref{ML_MLE} changes using a different stellar population model, stellar library, and full spectrum fitting method in Figure \ref{ML_FF}. Two BCGs now fall below the known correlations, but the scatter at higher mass-excess factor ($\alpha$) is now more pronounced. 

\begin{figure}
\centering
   \subfloat{\includegraphics[scale=0.42]{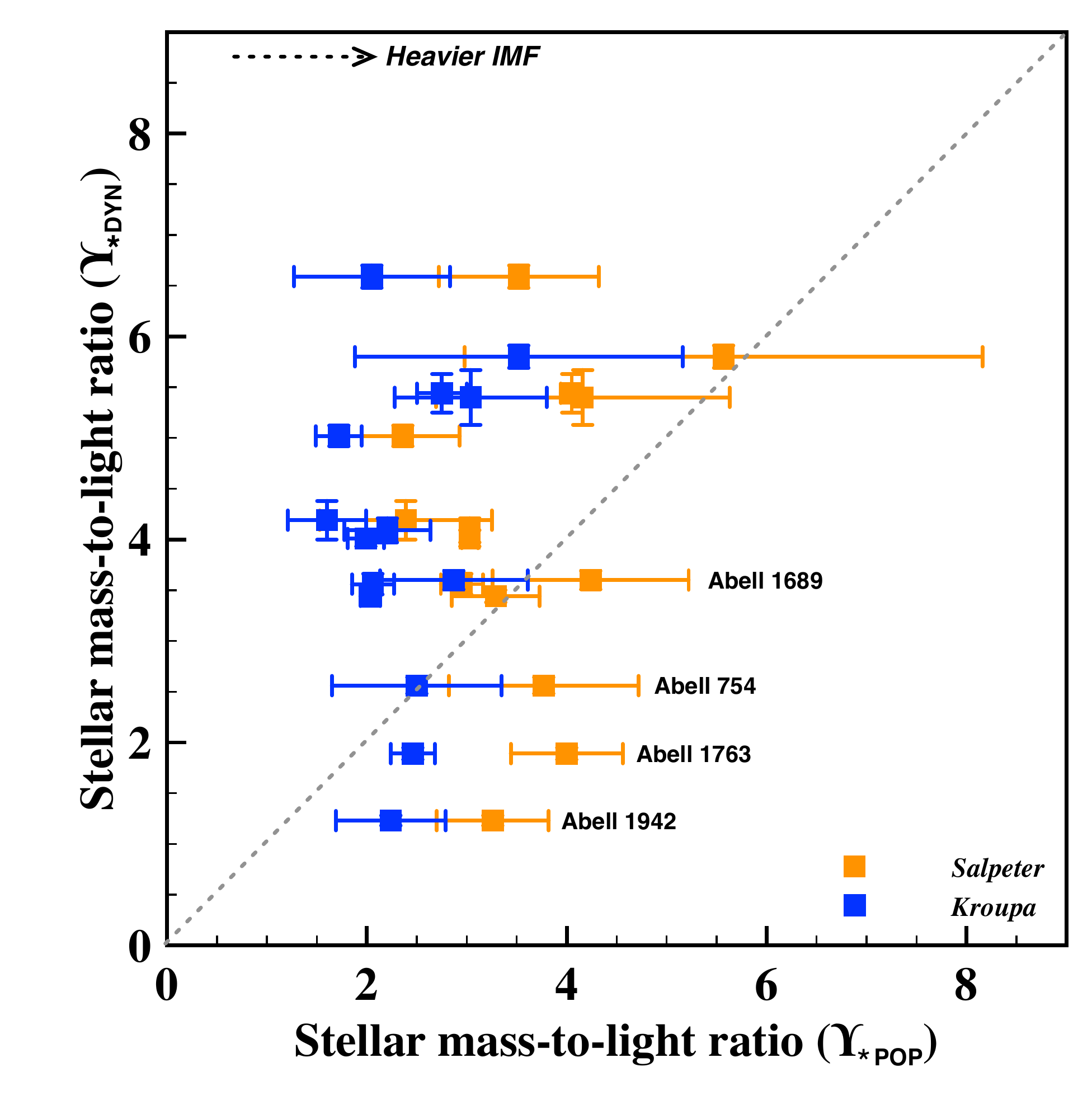}}
   \caption{$\Upsilon_{\star \rm DYN}$ vs $\Upsilon_{\star \rm POP}$ for Salpeter and Kroupa IMF (inner and outer apertures), where $\Upsilon_{\star \rm POP}$ is the average from ULySS/Vazdekis/MILES and FIREFLY/MaStar/E-MaStar.} 
\label{ML_ML_AVERAGE}
\end{figure}

This comparison (ULySS/Vazdekis/MILES and FIREFLY/MaStar/E-MaStar) indicates that realistic errors on $\Upsilon_{\star \rm POP}$ should be larger to include the systematic errors from using a different combination of stellar population model, library, and fitting method. Even though using a different stellar population analysis has a pronounced effect on the determination of $\Upsilon_{\star \rm POP}$, it does not eliminate the variety of IMFs necessary to describe the BCGs. In Figure \ref{ML_ML_AVERAGE}, we show that the average (and standard deviation) of the two different determinations of $\Upsilon_{\star \rm POP}$ still scatter above and below the 1-to-1 line. 
 
Our goal is not to do a comprehensive comparison between stellar population models, but to understand the effect different models have on determining $\Upsilon_{\star \rm POP}$. We further briefly directly compare ages and metallicities derived with six different combinations of fitting methods, stellar population models, and stellar libraries. We use: fitting methods ULySS \citep{Koleva2009}, and FIREFLY \citep{Wilkinson2017}; the stellar population models Vazdekis \citep{Vazdekis2010}, PEGASE-HR \citep{LeBorgne2004},  MaStar \citep{Maraston2020}, and M11 \citep{Maraston2011}; the stellar libraries MILES \citep{Sanchez2006}, ELODIE 3.1 \citep{Prugniel2001}, E-MaStar and Th-MaStar \citep{Yan2019, Maraston2020}.

\begin{enumerate}
\item ULySS + Vazdekis + MILES
\item ULySS + Pegase + ELODIE
\item FIREFLY + MaStar + E-MaStar
\item FIREFLY + MaStar + Th-MaStar
\item FIREFLY + M11 + MILES
\item FIREFLY + M11 + ELODIE
\end{enumerate}

We find that the average SSP-equivalent age and metallicity determined for six different combination fall within the average age and metallicity determined using ULySS/Vazdekis/MILES and FIREFLY/MaStar/E-MaStar, and no single combination can consistently derive an age and metallicity that would reconcile the $\Upsilon_{\star \rm POP}$ above as well as below the 1-to-1 line with $\Upsilon_{\star \rm DYN}$.

\section{The velocity dispersion within $\sigma_{e}$}
\label{apertures}

\begin{figure}
\centering
   \subfloat{\includegraphics[scale=0.34]{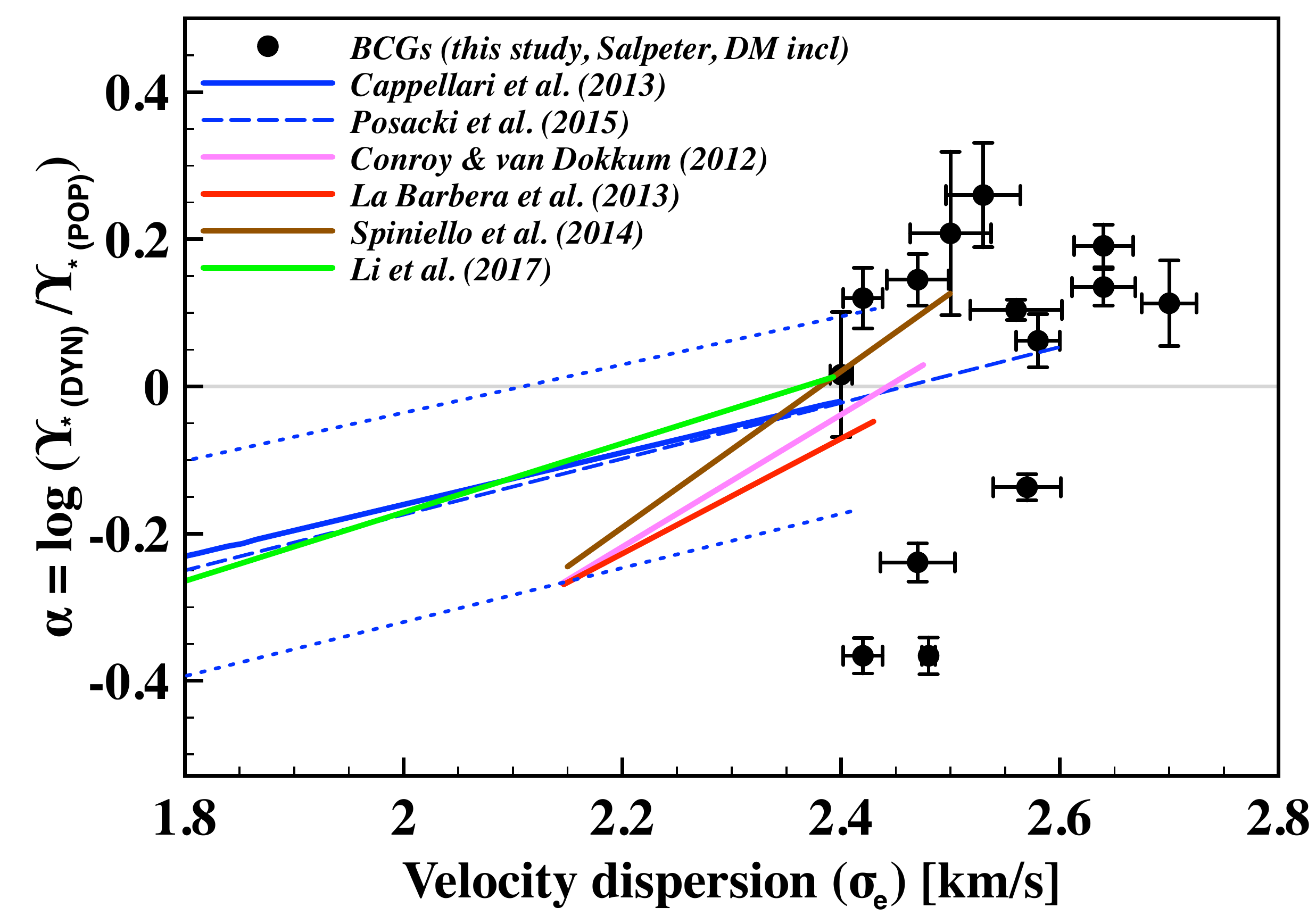}}
   \caption{We plot the mass-excess factor ($\alpha$) against velocity dispersion within the effective radius $\sigma_{e}$, and indicate the BCGs (from Figure \ref{ML_ML}) with black symbols. The legend is the same as in Figure \ref{ML_MLE}.} 
\label{ML_SigE}
\end{figure} 

We do an aperture correction using the velocity dispersion profiles measured in \citet{Loubser2018}, and $R_{e}$ measured in \citet{Loubser2020a}, to derive the velocity dispersion within the half-light radius, $\sigma_{e}$ and plot the aperture corrected plot in Appendix \ref{apertures}. However, our $\alpha$ measurement is limited to a 15 kpc aperture. Figure \ref{ML_SigE} shows that an aperture correction does not change our overall conclusion of substantial scatter in the IMF for the most massive galaxies.


\bsp	
\label{lastpage}
\end{document}